\documentclass[a4paper,11pt]{article}
\pdfoutput=1
\usepackage{jheppub}

\usepackage{amssymb,amsmath,mathtools,amsthm}
\usepackage{hyperref}
\usepackage{tikz}
\usetikzlibrary{calc,decorations.pathmorphing}
\usepackage{enumitem}
\usepackage{xcolor}
 \usepackage{subcaption}

\usepackage[T1]{fontenc}
\usepackage{lmodern}
\usepackage{microtype}
\usepackage{bm}
\usepackage{booktabs}
\usepackage{array}
\usepackage{placeins}
\usepackage{algorithm2e}
\usepackage{pgfplots}

\pgfplotsset{compat=1.18}
\hypersetup{colorlinks=true,linkcolor=red,citecolor=red,urlcolor=red}

\newtheorem{theorem}{Theorem}[section]

\theoremstyle{definition}

\theoremstyle{remark}

\newcommand{\ii}{\ensuremath{\mathrm{i}}} % imaginary unit
\newcommand{\vect}[1]{\bm{#1}}
\newcommand{\psd}{\succeq 0}
\newcommand{\R}{\mathbb{R}}

\newcommand{\floor}[1]{\left\lfloor #1\right\rfloor}

\theoremstyle{definition}

\theoremstyle{remark}

\usepackage{titlesec}
\usepackage{tcolorbox}

\newcommand{\cF}{\mathcal{F}}
\newcommand{\cU}{\mathcal{U}}
\newcommand{\cT}{\mathcal{T}}
\newcommand{\eps}{\varepsilon}
\newcommand{\dd}{\mathrm{d}}

\preprint{MPP-2026-140, TIF-UNIMI-2026-13}

\title{The Stieltjes Bootstrap for Feynman Integrals}

\author[a]{Hadrien Brochet,}
\author[a]{Maximilian Haensch,}
\author[a]{Johannes M. Henn,}
\author[b]{Benjamin Hollering,}
\author[a,d]{Prashanth Raman,}
\author[a]{Qinglin Yang,}
\author[c]{Simone Zoia}

\affiliation[a]{Max-Planck-Institut f\"ur Physik, Werner-Heisenberg-Institut,
Boltzmannstr.~8, 85748 Garching, Germany}

\affiliation[b]{Max Planck Institute for Mathematics in the Sciences, Inselstr. 22, 04103 Leipzig, Germany}

\affiliation[c]{INFN, Sezione di Milano, Via Celoria 16, 20133 Milan, Italy}
\affiliation[d]{Leung Center for Cosmology and Particle Astrophysics, Taipei 10617, Taiwan
}
\emailAdd{hbrochet@mpp.mpg.de}
\emailAdd{haensch@mpp.mpg.de}
\emailAdd{henn@mpp.mpg.de}
\emailAdd{benjamin.hollering@mis.mpg.de}
\emailAdd{praman@ntu.edu.tw}
\emailAdd{qlyang@mpp.mpg.de}
\emailAdd{simone.zoia@mi.infn.it}

\abstract{We prove that scalar Feynman integrals are Stieltjes functions in each
kinematic variable $-p_i\cdot p_j$, as well as in the particle masses, for
arbitrary loop order, arbitrary numbers of external legs, and arbitrary
internal masses, whenever a certain exponent condition is satisfied.
The proof rests on a canonical decomposition of the second Symanzik
polynomial with manifestly non-negative coefficients
indexed by spanning two-forests.
We present a dedicated analysis of the case in which the kinematic variables satisfy relations, as frequently happens in physical applications.
We argue that integration-by-parts relations allow the selection of
master-integral bases in which every element is a Stieltjes function.
The Stieltjes property guarantees the convergence of Pad\'e approximants
and provides determinantal (Hankel) constraints on the Taylor
coefficients of master integrals that go beyond previously known
complete-monotonicity bounds.  We develop a numerical strategy
that combines these Stieltjes constraints with differential equations
to determine the master integrals in the Euclidean region via 
semidefinite optimization.
We demonstrate this approach on the pedagogical example of the
six-dimensional massive box integral, and then apply it to
the calculation of previously unknown three-loop two-mass integrals needed for three-loop electroweak corrections to the $Z$- and Higgs-boson self-energies.}

\keywords{Scattering amplitudes, Feynman integrals, Pad\'e approximation,
Stieltjes functions, differential equations, semidefinite optimization}

\begin{document}

\maketitle
\flushbottom

%=====================================================================
\section{Introduction}
\label{sec:intro}
%=====================================================================

Feynman integrals are central objects in perturbative quantum field theory, appearing in the computation of virtually all observables of interest at particle colliders as well as in gravitational-wave physics~\cite{Travaglini:2022uwo,Huss:2025nlt,Buonanno:2022pgc}. In recent years, new mathematical structures have significantly advanced our ability to compute these integrals~\cite{Henn:2014qga,Abreu:2022mfk,Weinzierl:2022eaz}; nonetheless, the full evaluation of multi-loop Feynman integrals with multiple kinematic scales remains a formidable problem.  Since precision phenomenology demands ever-higher-order perturbative corrections, it is natural to ask: what structural properties of Feynman integrals can be exploited to make their evaluation more efficient and more reliable?

A variety of numerical approaches have been developed to address different aspects of this problem.  Methods based on sector decomposition~\cite{Binoth:2000ps} reorganize the Feynman-parameter integrand so that divergences factorize and can be subtracted, allowing for direct numerical integration.  Public implementations such as \textsc{pySecDec}~\cite{Borowka:2017idc} and \textsc{FIESTA}~\cite{Smirnov:2015mct} have been applied successfully to a wide range of multi-loop integrals.  The tropical Monte Carlo approach of Borinsky~\cite{Borinsky:2020rqs,Borinsky:2023jdv} provides an efficient sampling strategy in Feynman-parameter space; recent progress on subtraction schemes~\cite{Salvatori:2024subtrop,Giroux:2026tgd} is extending the scope of these methods to infrared-divergent integrals.

An important and complementary class of methods exploits the differential equations satisfied by Feynman integrals~\cite{Kotikov:1990kg,Bern:1993kr,Remiddi:1997ny,Gehrmann:1999as}.
Methods for solving these differential equations (DEs) include generalized series expansions~\cite{Moriello:2019yhu} implemented in publicly available tools~\cite{Hidding:2020ytt,Prisco:2025wqs,Armadillo:2022ugh}, numerical ordinary DE solvers~\cite{Czakon:2008zk}, and the ordinary-DE method of ref.~\cite{Mezzarobba:2025}, which comes with mathematically proven error bounds.
The auxiliary mass flow method, implemented in \textsc{AMFlow}~\cite{Liu:2017jxz,Liu:2022chg,Huang:2026rjb}, solves the boundary-value problem by introducing an auxiliary mass parameter, along which the integrals can be expanded in a region where the boundary conditions are trivially known. While \textsc{AMFlow} is powerful and broadly applicable, the introduction of the extra scale increases the size of the differential-equation system, which becomes the computational bottleneck for complex integral families.

Each of these methods has its strengths, and state-of-the-art calculations typically combine several of them. Still, the ever-growing need for precision and the associated computational cost remain ongoing challenges. This motivates the search for new structural properties that can either simplify the computation or provide independent cross-checks.

In this paper, we introduce and exploit one such structural property: the Stieltjes property of scalar Feynman integrals.  A function $f$ on $(-R,\infty)$, $R\geq0$, is a Stieltjes function if it admits an integral representation of the form
\begin{equation}
f(x) = \int_R^\infty \frac{\dd\mu(t)}{x+t}\,,
\qquad \dd\mu(t)\geq0\,,
\label{eq:stieltjes_intro}
\end{equation}
where $\mu$ is a non-negative measure supported on $[R,\infty)$. In ref.~\cite{Ditsch:2025dhp}, a subset of the present authors, together with S.~Ditsch, showed that scalar Feynman integrals are Stieltjes functions, or, more generally, completely monotone (CM) functions \cite{Henn:2024qwe}, when the propagator powers and dimension lie within a certain range, and this property was exploited for numerical evaluation.  However, the verification of the Stieltjes representation required a case-by-case decomposition of the second Symanzik polynomial, and the question of a general proof---especially for non-planar graphs and arbitrary numbers of external legs---was left open.

The first main result of this paper is a graph-theoretic proof of the Stieltjes
property that applies to any connected Feynman graph. For an arbitrary connected graph, the spanning-two-forest representation gives a canonical decomposition of the second Symanzik polynomial in the scalar products $-p_i\cdot p_j$, with coefficients that are manifestly non-negative throughout Feynman-parameter space.  Subject to convergence and an exponent condition, this proves that the corresponding
scalar integral is a Stieltjes function separately in each such variable and in each internal mass.
Physical specializations require additional care: on-shell constraints
or identifications between internal and external masses may remove the
positive orthant.  In section~\ref{sec:concrete-specialisations}, we formulate the positivity problem directly on the constrained kinematic space and
show how its solutions define adapted positive coordinates.  The argument underlying Theorem~\ref{thm:stieltjes} then applies
to each such coordinate, making it a Stieltjes variable.  For massless internal propagators, the same analysis can also establish that, in some cases, no
Euclidean region with non-empty interior exists on the physical constraint surface.

The Stieltjes property becomes especially useful for systems of differential equations when the master integrals are chosen to be Stieltjes functions.  We explain how higher-dimensional scalar integrals, dimensional recurrence relations and integration-by-parts (IBP) reduction~\cite{Tkachov:1981wb,Chetyrkin:1981qh,Laporta:2000dsw} can be used to search for such bases.

For a Stieltjes basis, the differential equations express all derivatives at a Euclidean point as linear combinations of the unknown master-integral values.
Positivity of the Hankel matrices built from Taylor coefficients leads to linear matrix inequalities that define a semidefinite program (SDP), whose solution gives upper and lower bounds on those values. We present the general setup of this new approach, and illustrate it on the
one-loop massive box. Moreover,
we apply this method to previously unknown three-loop integrals with two distinct masses.
The complete procedure is summarized as an algorithm in section~\ref{sec:diff-eq-sdp}.

The procedure applies at any fixed value of the dimensional regulator $\eps$, and the coefficients of
the Laurent expansion around $\eps=0$ follow from evaluations at several values of $\eps$ by standard
numerical reconstruction~\cite{Liu:2022chg,Zeng:2023jek}, as we show in section~\ref{sec:results}.

Semidefinite programming was previously applied to Feynman integrals in ref.~\cite{Zeng:2023jek},
where positivity of the Feynman-parameter integrand was combined with IBP relations, and in
ref.~\cite{Ditsch:2025dhp}, which exploited complete monotonicity.
Our
approach differs in two respects. First, we use the differential equations to generate, from a
finite set of master integrals, an unlimited number of constraints by iterating the derivative.
Second, the Stieltjes property yields Hankel and localizing-matrix inequalities that are
strictly stronger than the complete-monotonicity conditions exploited in ref.~\cite{Ditsch:2025dhp},
as we show explicitly in section~\ref{sec:diff-eq-sdp}.

The paper is organized as follows.    Section~\ref{sec:proof}
proves the Stieltjes property for scalar Feynman integrals,
treats constrained kinematic configurations, and explains the
construction of Stieltjes bases.
Section~\ref{sec:mi_bases} develops the Hankel/SDP bootstrap and applies it to the
one-loop massive box.  Section~\ref{sec:examples} contains
the application to the three-loop two-mass self-energy integral family.  We conclude
in section~\ref{sec:summary}.
Appendix~\ref{sec:stieltjes_review} collects the properties of Stieltjes functions
and Pad\'e approximants used in the text, appendix~\ref{app:matrices} gives the connection matrices of the box example, and appendix~\ref{app:euclidean_ladder} derives the Stieltjes domain of the three-loop family.

\section{Stieltjes Feynman integrals from Symanzik positivity}
\label{sec:proof}

This section establishes the property on which the rest of the paper rests: a scalar Feynman integral,
viewed as a function of any one of its kinematic invariants or internal masses, is a Stieltjes function in
the sense of eq.~\eqref{eq:stieltjes_intro}. The proof relies only on the
Feynman-parametric representation and the observation that the second Symanzik polynomial can be
decomposed with manifestly non-negative coefficients. In Section~\ref{sec:proof_stieltjes} we give this
proof for a generic configuration in which all invariants are independent. Physical applications almost
always impose relations among them, such as on-shell conditions or equal masses, and the positive
decomposition of the generic case can then degenerate; Section~\ref{sec:concrete-specialisations} shows
how to find the Stieltjes variables directly on the constrained kinematic space or to demonstrate that none exist. Section~\ref{sec:construction} turns the property into a practical tool by explaining how
to choose master integrals, using higher dimensions, propagators with higher power and dimensional recurrence, so that
every element of the basis is itself a Stieltjes function.

%=====================================================================

\subsection{Proof of the Stieltjes property}
\label{sec:proof_stieltjes}

We consider a connected Feynman graph $G$ with $L$ loops, $E$ internal
edges with masses $m_1,\ldots,m_E\geq 0$, and $n$ external legs carrying
momenta $p_1,\ldots,p_n$. The latter obey momentum conservation,
$\sum_{i=1}^n p_i=0$. Following ref.~\cite{Badger:2023xtl}, we use the
mostly-minus metric $\eta^{\mu\nu}=\mathrm{diag}(+,-,-,\ldots,-)$ and
the all-incoming convention. We introduce a Feynman parameter
$\alpha_e\geq 0$ for each internal edge and assume $D\geq n-1$ (or dimensional regularization), so that
there are no additional Gram-determinant constraints.
The Feynman-parametric
representation of a scalar integral reads (see e.g.\ ref.~\cite{Weinzierl:2022eaz})
\begin{equation}
  I = \frac{\Gamma(\lambda)}{\prod_e \Gamma(a_e)}
  \int_0^\infty \frac{\prod_e \dd \alpha_e\,\alpha_e^{a_e-1}}
  {\mathrm{GL}(1)}\;
  \frac{\cU^{\lambda-D/2}}{\cF^{\lambda}}\,,
\label{def:feynman_rep}
\end{equation}
where $a_e$ is the power of propagator $e$, $a=\sum_e a_e$, $\lambda=a-DL/2$, 
and $\mathrm{GL}(1)$ is the projective scaling group factored out to eliminate the overall scale redundancy of the Feynman parameters.
We refer to $\lambda$ as
the \emph{Stieltjes exponent}. The polynomials $\cU$ and $\cF$ are the first and
second Symanzik polynomials, respectively. The first Symanzik polynomial
is
\begin{equation}
  \cU = \sum_{T\in\cT_1(G)}\prod_{e\notin T}\alpha_e\,,
\label{eq:U_def}
\end{equation}
where $\cT_1(G)$ denotes the set of spanning trees of $G$. Since each
monomial is a product of non-negative Feynman parameters, $\cU\geq0$ is manifest. The second Symanzik polynomial encodes the kinematic dependence. It can
be expressed as a sum over spanning two-forests, that is, pairs of
disjoint trees that together span all vertices of $G$ and partition the
external legs into two complementary sets. For each forest, we denote by 
$I_F$ the set of external legs that does not contain the distinguished leg $n$.
Then
\begin{equation}
  \cF =
  -\sum_{F\in\cT_2(G)}
  \left(\sum_{i\in I_F}p_i\right)^2
  \prod_{e\notin F}\alpha_e
  +
  \cU\sum_{e=1}^E m_e^2\alpha_e\,,
\label{eq:F_def}
\end{equation}
where $\cT_2(G)$ denotes the set of spanning two-forests.

We first consider a generic kinematic configuration, in which all external legs $p_i$ are off-shell and no relations are imposed among the kinematic invariants and masses. We use the scalar products
\begin{equation}
  y_{ij}=-p_i\cdot p_j\,,
  \qquad 1\leq i\leq j\leq n-1\,,
\label{eq:yij_def}
\end{equation}
as kinematic variables.
These are the
$n(n-1)/2$ independent invariants of $n$-particle kinematics, apart from the
internal masses.
Together with the internal masses $m_e^2$, they define the set of
\emph{Stieltjes variables}
\begin{equation}
  \{\,y_{ij}:1\leq i\leq j\leq n-1\,\}
  \cup
  \{\,m_e^2:e=1,\ldots,E\,\}\,.
\label{eq:stieltjes_vars}
\end{equation}
Note that we have
\begin{equation}
    -\left(\sum_{i\in I_F}p_i\right)^2 =\sum_{i\in I_F}y_{ii}
    +2\!\sum_{\substack{i<j\\ i,j\in I_F}}\!y_{ij}\,.
\end{equation}
Thus, every Mandelstam invariant can be written as a linear combination, with non-negative integer
coefficients, of the scalar products $y_{ij}$, $1\leq i\leq j\leq n-1$.
Substituting this into the forest representation and collecting terms yields
\begin{equation}
  \cF
  =
  \sum_{1\leq i\leq j\leq n-1}
  y_{ij}\,\widetilde F_{ij}
  +
  \cU\sum_{e=1}^E m_e^2\alpha_e\,,
\label{eq:main_decomp}
\end{equation}
where
\begin{equation}
  \widetilde F_{ii}
  =
  \sum_{\substack{F\in\cT_2(G)\\ i\in I_F}}
  \prod_{e\notin F}\alpha_e\,,
  \qquad
  \widetilde F_{ij}
  =
  2\sum_{\substack{F\in\cT_2(G)\\ i,j\in I_F}}
  \prod_{e\notin F}\alpha_e\,,
  \quad i<j\,.
\label{eq:Fijs}
\end{equation}
Since each coefficient $\widetilde F_{ij}$ is manifestly non-negative, the polynomial $\cF$ is positive for
$\alpha_e>0$ when the independent Stieltjes variables lie in the positive
orthant. Therefore, for generic off-shell kinematics, every connected
graph has a non-empty Euclidean region, independently of its planarity and
loop order.

Given the decomposition in eq.~\eqref{eq:main_decomp}, the Stieltjes property
follows from a change of variables in the Feynman-parametric
representation~\eqref{def:feynman_rep}, as in
ref.~\cite{Ditsch:2025dhp}. We select one variable, say $y_{ij}$, and
hold all remaining kinematic variables fixed at values for which the
complementary polynomial defined below is non-negative. This condition is
satisfied, in particular, in the non-negative orthant. The second Symanzik polynomial
then takes the form
\begin{equation}
  \cF=y_{ij}A(\alpha)+B(\alpha)\,,
\label{eq:cF_split}
\end{equation}
where
\begin{equation}
  A(\alpha)=\widetilde F_{ij}\geq0\,,
  \qquad
  B(\alpha)=
  \sum_{(k,l)\neq(i,j)}y_{kl}\widetilde F_{kl}
  +\cU\sum_e m_e^2\alpha_e
  \geq0\,.
\end{equation}
The change of variables in ref.~\cite{Ditsch:2025dhp} then yields a
Stieltjes representation in $y_{ij}$ for $0<\lambda\leq1$.
Explicitly, we write $\mathcal{F}^{-\lambda} = A^{-\lambda} (y_{ij}+t)^{-\lambda}$ with $t=B/A \ge 0$ and use
\begin{align}
(y+t)^{-\lambda} =\frac{ \sin \pi \lambda}{\pi} \int_t^{\infty} \frac{(s-t)^{-\lambda} \dd s}{y+s} \,, \quad 0 <\lambda <1 \,,
\end{align}
with $y \equiv y_{ij}$(the case $\lambda=1$ being trivial); the Feynman-parameter integral then becomes an integral of $1/(y_{ij} + s)$ against a non-negative measure in $s$, which is the representation of eq.~\eqref{eq:stieltjes_intro}.

The same argument applies to each variable in turn. Hence $I$ is a Stieltjes function in every variable
of eq.~\eqref{eq:stieltjes_vars}.

We summarize the result in the following theorem.
\begin{tcolorbox}
\begin{theorem}[Stieltjes property]
\label{thm:stieltjes}
Let $I$ be a convergent scalar Feynman integral of the form given in
eq.~\eqref{def:feynman_rep}, with Stieltjes exponent
$0<\lambda\leq1$. Then $I$ is a Stieltjes function separately in each
of the variables in eq.~\eqref{eq:stieltjes_vars}, that is, in each
scalar product $y_{ij}$ and each internal mass $m_e^2$, with all
remaining variables held fixed such that the complementary polynomial
$B(\alpha)$ is non-negative. In particular, this condition holds in the
non-negative orthant of the remaining Stieltjes variables.
\end{theorem}
\end{tcolorbox}
For $\lambda>1$, the same argument establishes that the
integral is a generalized Stieltjes function of order $\lambda$
under the same convergence and positivity
assumptions~\cite{Karp:2011Stieltjes,Raman:2026LectureNotesPositivity}.

Several remarks are in order.
\begin{itemize}
    \item The planar analysis of ref.~\cite{Ditsch:2025dhp} was formulated in terms of
\begin{equation}
  X_{ij}=-\bigl(p_i+\cdots+p_{j-1}\bigr)^2\,.
\end{equation}
Each $X_{ij}$ is a non-negative linear combination of the variables
$y_{ab}=-p_a\cdot p_b$. The Stieltjes property in these variables is
therefore covered by the result above. The converse does not hold in
general, since expressing the $y_{ij}$ in terms of the $X_{ij}$
requires differences with coefficients of both signs.
\item Eliminating a different momentum through momentum conservation expresses the same invariants through
a different subset of the $y_{ij}$; this changes only the labeling of the
Stieltjes variables, not the conclusion. In practice, one may choose the
elimination that makes the relevant decomposition most transparent.
\item Physically relevant kinematic configurations
may lie on constraint surfaces that do not intersect the interior of the orthant used in the theorem.  For example, if all external legs are massless and on shell,
momentum conservation implies
\begin{equation}
    \sum_{1\leq i<j<n} y_{ij}=0,
\end{equation}
so the $y_{ij}$ cannot all be strictly positive.  Internal masses do not remove this obstruction, since they enter $\mathcal F$ only through the manifestly non-negative term
$\mathcal U\sum_e m_e^2\alpha_e$. Identifications between external and internal masses impose further constraints.  In the next
subsection, we determine Euclidean regions and adapted Stieltjes variables directly on such constrained kinematic spaces.

\end{itemize}

\subsection{Specialization to constrained kinematic configurations}
\label{sec:concrete-specialisations}

For a given Feynman graph $G$ with $E$ internal edges, let
$\mathcal F(\alpha;z)$ denote its second Symanzik polynomial, where
$\alpha=(\alpha_1,\ldots,\alpha_E)$ are the Feynman parameters and $z$ collectively
denotes the kinematic invariants and masses. The determination of the Euclidean region can
be formulated as the problem of finding all kinematic configurations for which
$\mathcal F$ is non-negative throughout the Feynman-parameter integration domain.
Introducing the standard simplex
\begin{equation}
 \Delta_{E-1}
 =\left\{\alpha\in\mathbb R_{\geq0}^{E}\ \middle|\
 \sum_{e=1}^{E}\alpha_e=1\right\},
 \label{eq:standard-simplex}
\end{equation}
we seek to determine the region
\begin{equation}
 \mathcal E_G
 =\left\{z\in\mathcal K\ \middle|\
 \mathcal F(\alpha;z)\geq0,
 \ \forall\alpha\in\Delta_{E-1}\right\},
 \label{eq:euclidean-region}
\end{equation}
where $\mathcal K$ denotes the kinematic parameter space, including any imposed relations
among the external invariants and masses. The non-negativity condition allows for zeros of
$\mathcal F$ on the boundary of the simplex, which are particularly relevant for massless
Feynman integrals.

Mathematically, eq.~\eqref{eq:euclidean-region} is a problem of polynomial positivity over
a semi-algebraic domain, or equivalently, the determination of the copositivity region of
the second Symanzik polynomial. Several general algorithms can be employed to address
this problem. Real quantifier-elimination methods, such as cylindrical algebraic
decomposition~\cite{Collins1975}, can in principle eliminate the Feynman parameters and yield an exact semi-algebraic description of $\mathcal E_G$ in terms of the kinematic variables. For
one-loop integrals, where $\mathcal F$ is quadratic in the Feynman parameters, the problem reduces to the copositivity of a symmetric matrix. More generally, positivity certificates, including those based on P\'olya's theorem~\cite{Sturmfels:2025}, can establish positivity at given kinematic
points or identify sufficient conditions defining Euclidean subregions.

For the applications below, the general positivity problem admits a
useful simplification.  We first impose all physical relations among
the kinematic invariants and masses, choose independent coordinates
on the resulting physical parameter space, and expand the specialized
second Symanzik polynomial in monomials of the Feynman parameters,
\begin{equation}
    \mathcal{F}(\alpha;z)
    =
    \sum_{\boldsymbol{\nu}}
    c_{\boldsymbol{\nu}}(z)\,
    \alpha^{\boldsymbol{\nu}}.
\end{equation}
Since the second Symanzik polynomial is linear in the kinematic
invariants and squared masses, the coefficients
$c_{\boldsymbol{\nu}}(z)$ are affine-linear functions of the chosen
kinematic coordinates.  Requiring
\begin{equation}
    c_{\boldsymbol{\nu}}(z)\geq 0
    \qquad\text{for all }\boldsymbol{\nu}
\end{equation}
therefore gives a set of linear inequalities that define a
coefficient-wise positive subregion of $\mathcal{E}_G$.  In the
language of the preceding paragraph, this is the simplest
coefficient-positivity certificate for the polynomial
$\mathcal{F}$.

For a general integral family, and in particular in the presence of
internal masses, coefficient-wise non-negativity is a sufficient but
not a necessary condition for $\mathcal F\geq0$ on the simplex
\cite{Sturmfels:2025}.  Negative monomial coefficients may still combine into
positive polynomials, so failure of this criterion does not by itself
imply that the Euclidean region is empty.  The situation simplifies
when all internal propagators are massless.  The spanning-forest
representation then implies that $\mathcal F$ is homogeneous of
degree $L+1$ and square-free in the Feynman parameters.  The
coefficient of any monomial can therefore be isolated by setting to
zero all Feynman parameters outside its support.  Consequently, in
the massless case, $\mathcal F\geq0$ on the full simplex if and only
if all collected monomial coefficients are non-negative.  The
coefficient inequalities then determine the full Euclidean region
rather than merely a sufficient subregion.

Once a non-empty region has been found, we introduce adapted coordinates $x_a>0$ that parametrize it.
The change of variables is chosen such that the specialized Symanzik
polynomial takes the form
\begin{equation}
    \mathcal{F}(\alpha;x)
    =
    A_0(\alpha)
    +
    \sum_a x_a A_a(\alpha) \,,
    \qquad
    A_0(\alpha)\geq0 \,,
    \qquad
    A_a(\alpha)\geq0 \,.
\end{equation}
For any fixed $a$, holding the remaining $x_b$ fixed in their
positive region gives
\begin{equation}
    \mathcal{F}
    =
    x_a A_a(\alpha)
    +
    \left[
        A_0(\alpha)
        +
        \sum_{b\neq a}x_b A_b(\alpha)
    \right] \,,
\end{equation}
where both the coefficient of $x_a$ and the bracketed remainder are
non-negative throughout the integration domain.  The argument used
in the proof of Theorem~\ref{thm:stieltjes} therefore applies: subject to the
same convergence and Stieltjes-exponent conditions, the corresponding
Feynman integral is a Stieltjes function separately in every adapted
variable $x_a$.  The adapted variables are thus not merely
coordinates on the Euclidean region; they are precisely the Stieltjes
variables relevant for the subsequent construction.

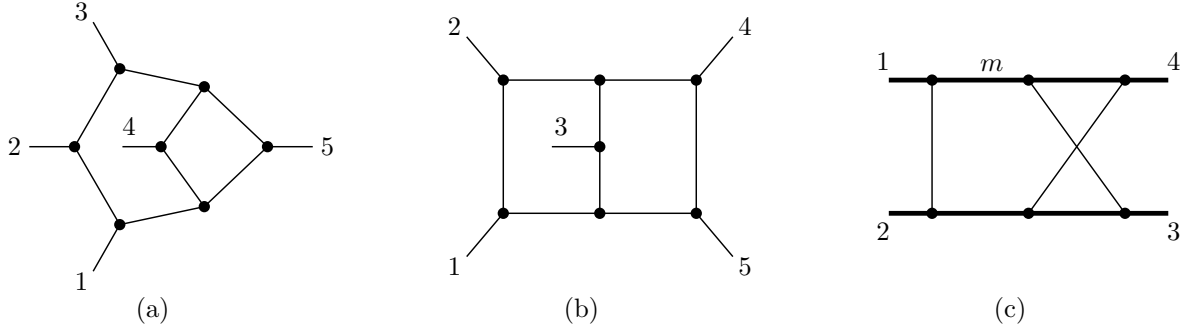
\begin{figure}[t]
\centering

% =========================================================
% (a) Non-planar hexa-box
% =========================================================
\begin{subfigure}[b]{0.25\textwidth}
\centering
\begin{tikzpicture}[
  scale=0.88,
  every node/.style={font=\small},
  vertex/.style={circle,fill=black,inner sep=1.5pt},
  massive/.style={line width=1.8pt},
  massless/.style={line width=0.55pt},
  ordinary/.style={line width=0.55pt},
  external/.style={line width=0.55pt}
]

  % Common bounding box: centred at y=0.55
  \path[use as bounding box]
    (-0.95,-1.40) rectangle (3.85,2.50);

  \coordinate (L) at (0,0.55);
  \coordinate (A) at (0.68,1.72);
  \coordinate (B) at (1.95,1.45);
  \coordinate (C) at (2.90,0.55);
  \coordinate (D) at (1.95,-0.35);
  \coordinate (E) at (0.68,-0.62);
  \coordinate (M) at (1.30,0.55);

  % Internal propagators
  \draw[ordinary] (L)--(A)--(B)--(C)--(D)--(E)--(L);
  \draw[ordinary] (M)--(B);
  \draw[ordinary] (M)--(D);

  % External legs
  \draw[external] (L)--+(-0.68,0);
  \draw[external] (A)--+(-0.40,0.70);
  \draw[external] (E)--+(-0.40,-0.70);
  \draw[external] (C)--+(0.68,0);
  \draw[external] (M)--+(-0.58,0);

  % Momentum labels
  \node[left]       at ($(L)+(-0.65,0)$)     {$2$};
  \node[above left] at ($(A)+(-0.32,0.58)$)  {$3$};
  \node[below left] at ($(E)+(-0.32,-0.58)$) {$1$};
  \node[right]      at ($(C)+(0.65,0)$)      {$5$};
  \node[above]      at ($(M)+(-0.48,0.02)$)  {$4$};

  % Vertices
  \foreach \v in {L,A,B,C,D,E,M}
    \node[vertex] at (\v) {};

\end{tikzpicture}
\caption{}
\label{fig:hexa-box}
\end{subfigure}\hfill%
% =========================================================
% (b) Double pentagon
% =========================================================
\begin{subfigure}[b]{0.25\textwidth}
\centering
\begin{tikzpicture}[
  scale=0.88,
  every node/.style={font=\small},
  vertex/.style={circle,fill=black,inner sep=1.5pt},
  massive/.style={line width=1.8pt},
  massless/.style={line width=0.55pt},
  ordinary/.style={line width=0.55pt},
  external/.style={line width=0.55pt}
]

  % Common bounding box: centred at y=0.55
  \path[use as bounding box]
    (-0.95,-1.40) rectangle (3.85,2.50);

  \coordinate (A)  at (0,-0.45);
  \coordinate (B)  at (0,1.55);
  \coordinate (M1) at (1.45,-0.45);
  \coordinate (X)  at (1.45,0.55);
  \coordinate (M2) at (1.45,1.55);
  \coordinate (C)  at (2.90,1.55);
  \coordinate (D)  at (2.90,-0.45);

  % Internal propagators
  \draw[ordinary] (A)--(B);
  \draw[ordinary] (B)--(M2)--(C);
  \draw[ordinary] (C)--(D);
  \draw[ordinary] (D)--(M1)--(A);
  \draw[ordinary] (M1)--(X)--(M2);

  % External legs
  \draw[external] (A)--+(-0.55,-0.65);
  \draw[external] (B)--+(-0.55,0.65);
  \draw[external] (C)--+(0.55,0.65);
  \draw[external] (D)--+(0.55,-0.65);
  \draw[external] (X)--+(-0.72,0);

  % Momentum labels
  \node[below left]  at ($(A)+(-0.48,-0.52)$) {$1$};
  \node[above left]  at ($(B)+(-0.48,0.52)$)  {$2$};
  \node[above right] at ($(C)+(0.48,0.52)$)   {$4$};
  \node[below right] at ($(D)+(0.48,-0.52)$)  {$5$};
  \node[above]       at ($(X)+(-0.58,0.02)$)  {$3$};

  % Vertices
  \foreach \v in {A,B,M1,X,M2,C,D}
    \node[vertex] at (\v) {};

\end{tikzpicture}
\caption{}
\label{fig:double-pentagon}
\end{subfigure}\hfill%
% =========================================================
% (c) Non-planar Bhabha integral
% =========================================================
\begin{subfigure}[b]{0.25\textwidth}
\centering
\begin{tikzpicture}[
  scale=0.88,
  every node/.style={font=\small},
  vertex/.style={circle,fill=black,inner sep=1.5pt},
  massive/.style={line width=1.8pt},
  massless/.style={line width=0.55pt},
  ordinary/.style={line width=0.55pt},
  external/.style={line width=0.55pt}
]

  % Common bounding box: centred at y=0.55
  \path[use as bounding box]
    (-0.95,-1.40) rectangle (3.85,2.50);

  \coordinate (A) at (0,1.55);
  \coordinate (B) at (1.45,1.55);
  \coordinate (C) at (2.90,1.55);

  \coordinate (D) at (0,-0.45);
  \coordinate (E) at (1.45,-0.45);
  \coordinate (F) at (2.90,-0.45);

  % Massive fermion lines
  \draw[massive] (A)--(B)--(C);
  \draw[massive] (D)--(E)--(F);

  % Massless photon lines
  \draw[massless] (A)--(D);
  \draw[massless] (B)--(F);
  \draw[massless] (E)--(C);

  % External legs
  \draw[massive] (A)--+(-0.65,0);
  \draw[massive] (D)--+(-0.65,0);
  \draw[massive] (C)--+(0.65,0);
  \draw[massive] (F)--+(0.65,0);

  % External-momentum labels
  \node[above left]  at ($(A)+(-0.48,0)$) {$1$};
  \node[below left]  at ($(D)+(-0.48,0)$) {$2$};
  \node[above right] at ($(C)+(0.48,0)$)  {$4$};
  \node[below right] at ($(F)+(0.48,0)$)  {$3$};

  % Mass label
  \node[align=left] at (0.9,1.78) {$m$};

  % Vertices
  \foreach \v in {A,B,C,D,E,F}
    \node[vertex] at (\v) {};

\end{tikzpicture}
\caption{}
\label{fig:bhabha-double-box}
\end{subfigure}

\caption{
(a) Five-point non-planar hexa-box;
(b) five-point double-pentagon;
(c) non-planar Bhabha integral.
}
\label{fig:non-planar-families}
\end{figure}

We now illustrate the coefficient-wise analysis with the three representative
examples in figure~\ref{fig:non-planar-families}.  The massless non-planar hexa-box in figure~\ref{fig:non-planar-families}(\subref*{fig:hexa-box})
 provides a case in which the
coefficient-wise conditions determine a non-empty Euclidean region and lead
directly to adapted Stieltjes variables.  For the massless double-pentagon in figure~\ref{fig:non-planar-families}(\subref*{fig:double-pentagon}),
the same conditions are incompatible with the physical on-shell constraint
and therefore prove the absence of a Euclidean region with non-empty
interior.  Finally, the equal-mass non-planar Bhabha integral in figure~\ref{fig:non-planar-families}(\subref*{fig:bhabha-double-box}) illustrates a
massive case in which the coefficient-wise criterion is too restrictive:
although it produces only a degenerate region, a nontrivial positive
subregion can still be exhibited by reorganizing the Symanzik polynomial
into positive quadratic forms.

\smallskip

{\it \underline{1.\ Massless non-planar hexa-box, cf.\ figure~\ref{fig:non-planar-families}(\subref*{fig:hexa-box}).}} The second Symanzik polynomial is
\begin{equation}
\begin{aligned}
\mathcal{F}=2\Bigl[&
 y_{12}\bigl(
   \alpha_1\alpha_3\alpha_{5678}
   +\alpha_3\alpha_5\alpha_7
 \bigr)
+y_{45}\bigl(
   \alpha_1\alpha_4\alpha_{5678}
   +\alpha_1\alpha_6\alpha_8
   +\alpha_4\alpha_5\alpha_7
 \bigr)
+y_{14}\alpha_2\alpha_5\alpha_8
\\
&+y_{23}\bigl(
   \alpha_2\alpha_4\alpha_{5678}
   +\alpha_2\alpha_6\alpha_8
 \bigr)
+y_{51}\alpha_2\alpha_6\alpha_7
+y_{35}\alpha_3\alpha_5\alpha_8
+y_{34}\alpha_3\alpha_6\alpha_7
\Bigr] \,.
\end{aligned}
\label{eq:hexabox_symanzik}
\end{equation}
Following the general strategy described above, we collect equal monomials
in the Feynman parameters and impose positivity on their kinematic
coefficients.  The resulting inequalities define the region
\begin{equation}
  y_{12}\geq y_{34}+y_{45} \,,\qquad
  y_{23}\geq y_{45}+y_{51} \,,\qquad
  y_{34},y_{45},y_{51}\geq 0 \,.
\label{eq:hexabox_v_region}
\end{equation}
Since all internal propagators are massless, coefficient-wise positivity is
also necessary for positivity of the second Symanzik polynomial on the full
Feynman-parameter simplex.  Therefore, these inequalities determine the
interior of the Euclidean region rather than merely a sufficient
subregion.
Introducing the variables
\begin{equation}
\{x_i\}_{i=1,\ldots,5}
=
\{y_{34},\ y_{45},\ y_{51},\
  y_{12}-y_{34}-y_{45},\
  y_{23}-y_{45}-y_{51}\} \,,
\label{eq:hexabox_euclidean_variables}
\end{equation}
the second Symanzik polynomial becomes
\begin{equation}
\begin{aligned}
  &\cF
  ={}2[
  (x_1{+}x_2{+}x_4)
  \bigl(
    \alpha_1\alpha_3\alpha_{5678}
    {+}\alpha_3\alpha_5\alpha_7
  \bigr){+}
  x_2
  \bigl(
    \alpha_1\alpha_4\alpha_{5678}
    {+}\alpha_1\alpha_6\alpha_8
    {+}\alpha_4\alpha_5\alpha_7
  \bigr)
\\
&{+}
  (x_2{+}x_3{+}x_5)
  \bigl(
    \alpha_2\alpha_4\alpha_{5678}
    {+}\alpha_2\alpha_6\alpha_8
  \bigr){+}
  x_5\alpha_2\alpha_5\alpha_8
  {+}x_3\alpha_2\alpha_6\alpha_7
  {+}x_4\alpha_3\alpha_5\alpha_8
  {+}x_1\alpha_3\alpha_6\alpha_7]\,.
\end{aligned}
\label{eq:hexabox_physical_region_decomp}
\end{equation}
This is a coefficient-wise non-negative decomposition of
$\mathcal{F}$. Hence the Euclidean region is the positive orthant 
\begin{equation}
  \mathcal{E}
  =
  \{x_1\geq 0,\ x_2\geq0,\ x_3\geq0,\ x_4\geq0,\ x_5\geq0\} \,.
\label{eq:hexabox_x_region}
\end{equation}
which agrees with
ref.~\cite{Chicherin:2018mue}.
The variables $x_i$ therefore serve both as positive coordinates on the
Euclidean region and as adapted Stieltjes variables.  Subject to the
convergence and exponent conditions of Theorem~\ref{thm:stieltjes}, the
corresponding integrals are Stieltjes functions separately in each $x_i$.

\smallskip
{\it \underline{2.\ Massless non-planar double-pentagon, cf.\ figure~\ref{fig:non-planar-families}(\subref*{fig:double-pentagon}).}} Applying the same coefficient-wise analysis to this example gives the necessary inequalities
\begin{equation}
  y_{13}\geq0 \,,\qquad
  y_{23}\geq0 \,,\qquad
  y_{34}\geq0 \,,\qquad
  y_{12}+y_{14}+y_{24}\geq0 \,.
\label{eq:double_pentagon_inequalities}
\end{equation}
Their sum, however, is fixed by the physical on-shell constraint,
\begin{equation}
  y_{12}+y_{13}+y_{14}+y_{23}+y_{24}+y_{34}=0 \,.
\label{eq:double_pentagon_constraint}
\end{equation}
The inequalities therefore imply
\begin{equation}
     y_{13}=0 \,,\qquad
  y_{23}=0 \,,\qquad
  y_{34}=0 \,,\qquad
  y_{12}+y_{14}+y_{24}=0
\end{equation}
on the physical massless kinematic surface.  Because the internal propagators are
massless, the coefficient-wise criterion is exact in this example.  Its
failure is consequently not merely a failure of a sufficient positivity
certificate: it proves that the physical massless double-pentagon has no
Euclidean region with non-empty interior.  Accordingly, no set of adapted
positive Stieltjes variables can be obtained on the physical surface for
this family.
A standard way around this obstruction is to keep one external leg off shell, which restores a positive orthant, and to take the on-shell limit at the end, as in the differential-equation and Mellin–Barnes treatments of massless non-planar integrals in refs.~\cite{Henn:2014qga,Tausk:1999vh}.

\smallskip
{\it \underline{3.\ Equal-mass non-planar Bhabha integral, cf.\ figure~\ref{fig:non-planar-families}(\subref*{fig:bhabha-double-box}).}} Here the kinematic variables satisfy
\begin{equation}
  y_{12}+y_{13}+y_{23}=m^2.
\label{eq:bhabha_constraint}
\end{equation}
After eliminating $y_{23}$, the second Symanzik polynomial takes the form
\begin{equation}
  \mathcal{F}_{\mathrm{NP}}
  =
  y_{12}C_{12}
  +y_{13}C_{13}
  +m^2 C_{m^2} \,,
\label{eq:bhabha_specialized_F}
\end{equation}
where
\begin{equation}
\begin{aligned}
C_{12}
&=
2\Bigl(
 \alpha_2\alpha_5\alpha_{34}
+\alpha_3\alpha_5\alpha_{467}
+\alpha_3\alpha_6\alpha_7
-\alpha_1\alpha_2\alpha_7
\Bigr) \,,
\\
C_{13}
&=
2\alpha_1
\Bigl(
 \alpha_4\alpha_6-\alpha_2\alpha_7
\Bigr) \,,
\\
C_{m^2}
&=
\alpha_1\bigl(\alpha_4^2+\alpha_6^2\bigr)
+\alpha_2\bigl(\alpha_{34}^2+\alpha_5^2\bigr)
+\alpha_3^2\alpha_{467}
+\alpha_3\alpha_{46}^2
\\
&\quad
+\alpha_4^2\alpha_{56}
+\alpha_4\alpha_{56}^2
+\alpha_5^2\alpha_{67}
+\alpha_5\alpha_6\alpha_{67}
+\alpha_6\alpha_7\alpha_{56} \,.
\end{aligned}
\label{eq:bhabha_coefficients}
\end{equation}
We first apply the coefficient-wise criterion directly to this
specialized polynomial.  After expanding the expressions above and
collecting equal monomials in the Feynman parameters, coefficient-wise
non-negativity requires
\begin{equation}
  m^2\geq0 \,,\qquad
  y_{12}\geq0 \,,\qquad
  y_{13}\geq0 \,,\qquad
  y_{12}+y_{13}\leq0 \,.
\label{eq:bhabha_coefficientwise_conditions}
\end{equation}
These conditions imply
\begin{equation}
  y_{12}=y_{13}=0 \,,\qquad m^2\geq0 \,,
\label{eq:bhabha_degenerate_region}
\end{equation}
and hence do not define a full-dimensional Euclidean region.

Unlike in the preceding massless example, this result does not prove
that the Bhabha family has no nontrivial Euclidean region.  In the presence
of internal masses, coefficient-wise non-negativity is sufficient but not
necessary for positivity of the second Symanzik polynomial.  In particular,
monomials with negative coefficients may combine with other terms into
positive quadratic forms.  A complete quantifier-elimination or
copositivity analysis could in principle determine the full
semi-algebraic Euclidean region.  Here we do not perform this more general
elimination.  Instead, we exhibit explicitly a nontrivial positive cone
that is missed by the direct coefficient-wise test.

We introduce the variables
\begin{equation}
  m^2=u+v+w \,,\qquad
  y_{12}=-u+v-w \,,\qquad
  y_{13}=-u-v+w \,.
\label{eq:bhabha_adapted_variables}
\end{equation}
Substituting these relations gives
\begin{equation}
  \mathcal{F}_{\mathrm{NP}}
  =
  uA_u+vA_v+wA_w \,,
\label{eq:bhabha_positive_decomposition}
\end{equation}
where
\begin{align}
    A_u={}&
    \alpha_1\left[
        (\alpha_4{-}\alpha_6)^2
        +4\alpha_2\alpha_7
    \right]
    +\alpha_2(\alpha_{34}{-}\alpha_5)^2
    +\alpha_7(\alpha_3{-}\alpha_5{-}\alpha_6)^2
    +\alpha_4^2\alpha_{356}\nonumber\\
    &+
    \alpha_{46}
        (\alpha_3{-}\alpha_5)^2
        +2\alpha_4\alpha_6\alpha_{35}
        +\alpha_{345}\alpha_6^2 \,,\\
    A_v={}&
    \alpha_1(\alpha_4-\alpha_6)^2
    +\alpha_2(\alpha_{34}+\alpha_5)^2
    +\alpha_3^2\alpha_{467}
    +\alpha_3\alpha_{46}^2+
    \alpha_4^2\alpha_{56}
    +\alpha_4\alpha_{56}^2
    \nonumber\\
    &
    +\alpha_5^2\alpha_{67}
    +\alpha_5\alpha_6\alpha_{67}
    +\alpha_6\alpha_7\alpha_{56}+2\alpha_3\alpha_5\alpha_{467}
    +2\alpha_3\alpha_6\alpha_7 \,,
    \\
    A_w={}&
    \alpha_1\alpha_{46}^2
    +\alpha_2(\alpha_{34}-\alpha_5)^2
    +\alpha_7(\alpha_3{-}\alpha_5{-}\alpha_6)^2
    +\alpha_4^2\alpha_{356}\nonumber\\
    &+
    \alpha_{46}
        (\alpha_3{-}\alpha_5)^2
        +2\alpha_4\alpha_6\alpha_{35}
        +\alpha_{345}\alpha_6^2 \,.
\end{align}
All terms in these expressions are non-negative for $\alpha_e\geq0$.
Consequently,
\begin{equation}
  \mathcal{E}_{\mathrm{NP}}^{\mathrm{sub}}
  =
  \{u\geq0,\ v\geq0,\ w\geq0\}
\label{eq:bhabha_positive_region}
\end{equation}
is a non-empty Euclidean subregion of the equal-mass Bhabha family.  The
superscript emphasizes that this construction gives a sufficient positive
region rather than the complete copositivity region.  Moreover, since
$\mathcal{F}_{\mathrm{NP}}=uA_u+vA_v+wA_w$ with
$A_u,A_v,A_w\geq0$, the argument of
Theorem~\ref{thm:stieltjes} applies separately to $u$, $v$, and $w$.
They therefore provide adapted Stieltjes variables within this subregion.

\smallskip

To summarize, the three examples illustrate the possible outcomes of the coefficient-wise analysis. For massless internal propagators the criterion is exact. As illustrated above, it produces adapted Stieltjes variables for the hexa-box and proves the absence of a Euclidean region with non-empty interior for the double pentagon. For the equal-mass Bhabha integral the criterion is only sufficient and yields a degenerate region, whereas a reorganization into positive quadratic forms exposes a non-empty subregion with adapted variables $u, v, w$. Coefficient-wise positivity is thus a cheap certificate when it applies, while massive non-planar families may require more general positivity or elimination methods.

\subsection{Construction of Stieltjes bases}
\label{sec:construction}

Theorem~\ref{thm:stieltjes} tells us that constructing Stieltjes integrals within an integral family amounts to finding sets of integers $\{a_1,\ldots,a_E,D_0\}$, $D=D_0-2\eps$, such that the Stieltjes exponent
\begin{equation}
    \lambda=\sum_{e=1}^E a_e-\tfrac{L D}{2}
\end{equation}
satisfies $0<\lambda\le 1$.
In doing so we have the freedom to choose \emph{different} spacetime dimensions $D_0$. For example, consider the one-loop box integral with unit propagator powers in $D_0=6$; this leads to $\lambda =1+\eps$, and hence it is Stieltjes for $-1<\eps \le 0$.
Such higher-dimensional integrals are related to the standard four-dimensional ones by dimensional recurrence relations~\cite{Tarasov:1996br}.

Let us assume that we are interested in integrals in $D_0=4$ dimensions.
Since dimensional
recurrence relations shift the dimension in steps of two, the relevant higher-dimensional
integrals carry even integer $D_0$.
Since $\lambda$ is an integer at $\eps = 0$,
the Stieltjes condition forces $\lambda=1$.
In dimensional regularization, $D=D_0-2\eps$ with
$\eps<0$ and $L |\eps|<1$, and the Stieltjes exponent becomes $\lambda=1+L\eps\leq1$.

Note that our Stieltjes proof assumes convergent integrals, and so the choice of $D$ and propagator powers is additionally constrained by ultraviolet (UV) finiteness: for
every UV subgraph $\gamma\subseteq G$, the usual parametric convergence condition is
\begin{equation}
 a_\gamma-\frac{L_\gamma D_0}{2}>0\,,
 \qquad
 a_\gamma:=\sum_{e\in\gamma}a_e\,,
 \label{eq:subgraph-finiteness}
\end{equation}
where $L_\gamma$ is the loop number of $\gamma$. Together with the condition $0<\lambda\leq1$ on the overall Stieltjes exponent,
these are the constraints we impose when selecting basis integrals.
Infrared finiteness imposes the analogous condition on the contracted (quotient) graphs; for the massive integral families considered in this paper only the ultraviolet condition is relevant.
The conditions above are linear in the integer variables $a_e$ and $D_0$. One can
therefore enumerate admissible scalar integrals by solving an integer programming problem.
Since the spacetime dimension $D_0$ can be taken to be arbitrarily large, the set of
admissible Stieltjes integrals is infinite (before considering IBP relations).

In all cases we have investigated, we found that
the span of these integrals under IBP reduction covers
the master-integral space of a given integral family.
In practice, we find such a basis by writing down a list of Stieltjes integrals with some practical bounds on the dimension and propagator powers. Note that this completeness statement is global: it concerns the span of the full IBP family and does not imply that every individual subsector contains a finite scalar Stieltjes representative in an even dimension.
Explicit Stieltjes bases are given below for the one-loop massive box
in section~\ref{sec:massive-box} and for the three-loop integral family
in section~\ref{sec:examples}.

%=====================================================================
\section{Numerical Stieltjes bootstrap from Hankel matrices}
\label{sec:mi_bases}
%=====================================================================

The moments of a Stieltjes function $f(x)$ must form positive semidefinite Hankel matrices, which provides nonlinear, namely determinantal, constraints. We now show how these moments may be obtained from the derivatives of $f$ and then use this to develop a method for bounding the value of $f(x_0)$ at a given point $x_0$. This method uses the fact that $f(x_0)$ must lie in a descending chain of spectrahedra and thus its value can be accurately approximated with semidefinite programming.

\subsection{Hankel matrix constraints for Stieltjes functions}
\label{app:hankel_stieltjes}

Let $\mu$ be a positive measure supported on $[R,\infty)$ and let
\begin{equation}
  f(x)=\int_R^\infty \frac{\dd \mu(t)}{x+t}
\end{equation}
be a Stieltjes function that is analytic off the cut $(-\infty,-R]$. Then, at a point $x_0>-R$, the Taylor coefficients of $f$ at $x_0$, with alternating signs stripped, are given by the moments of the measure $\mu$, which are defined as
\begin{equation}
  a_k=\frac{(-1)^k}{k!}f^{(k)}(x_0)
  =\int_R^\infty\frac{\dd \mu(t)}{(x_0+t)^{k+1}} \,,
  \qquad k=0,1,2,\ldots
\end{equation}
Our goal is to use positivity constraints on this sequence to approximate $f(x_0)$.

Let us illustrate this point by a simple example. Consider the function
\begin{equation}
  f(x)=\int_1^2\frac{\dd t}{x+t}=\log\frac{x+2}{x+1} \,.
\end{equation}
At the point $x_0=0$, the coefficients are simply
\begin{equation}
  a_0=\log 2,\qquad a_k=\frac{1-2^{-k}}{k} \,,\quad k\geq 1\,.
\end{equation}
Since $\dd\mu = \dd t$ on $[1,2]$ is a positive measure, the Hankel matrix $H_0^{(N)}=(a_{i+j})_{i,j=0}^{\floor{N/2}}$ (defined in general in eq.~\eqref{eq:Hankels} below) must be positive semidefinite for all $N$, i.e.\ every principal minor of $H_0^{(N)}$ must be non-negative. 
Assuming that we can compute the moments $a_k$ for $k\geq1$ at $x_0$,
positive semidefiniteness (PSD) constrains the value of $f(x_0)=a_0$. 
At $N=1$, PSD simply implies that $f(x_0) \ge 0$.
At $N=2$, we have that
\[
H_0^{(2)}=
\begin{pmatrix}a_0&a_1\\a_1&a_2\end{pmatrix}
=\begin{pmatrix}a_0 &\tfrac12\\[1mm]\tfrac12&\tfrac38\end{pmatrix}\psd
\quad\Longleftrightarrow\quad
f(x_0)\geq \frac23 \,.
\]
As $N$ increases, $H_0^{(N)} \psd $ gives an increasingly tight lower bound on $f(x_0)=\log2$, while only requiring that we can compute $a_k$ for $k\geq1$ at $x_0$.

We now describe a minimal family of PSD constraints which suffice to determine $\mu$ and hence $f(x)$. For any $c_i \in \mathbb{R}$, observe that
\begin{equation}
  p(t) \coloneq \sum_{i=0}^k\frac{c_i}{(x_0+t)^i}
\end{equation}
satisfies $p(t)^2\geq0$. Thus for any nonnegative function $h(t)\geq0$ on $[R, \infty)$ it holds that
\begin{equation}
0\leq\int_R^\infty\frac{\dd\mu(t)}{x_0+t}h(t)p(t)^2
=\sum_{i,j=0}^kc_i
\left(\int_R^\infty\frac{\dd\mu(t)h(t)}{(x_0+t)^{1+i+j}}\right)c_j \,.
\end{equation}
Since this inequality holds for all $c\in\R^{k+1}$, we see that the Hankel matrix with $(i,j)$ entry
\begin{equation}
  \int_R^\infty\frac{\dd\mu(t)h(t)}{(x_0+t)^{1+i+j}}
\end{equation}
is indeed positive semidefinite. Choosing different nonnegative functions $h(t)$ yields different constraints on $f(x_0)$. Possible choices are, for example,
\begin{equation}
h_1(t)=1\,,\qquad h_2(t)=\frac1{x_0+t}\,,\qquad
h_3(t)=\frac{t-R}{x_0+t}\,,\qquad h_4(t)=\frac{t-R}{(x_0+t)^2}\,,
\end{equation}
which correspond to the PSD constraints
\begin{equation}\label{eq:fourHankelineq}
\begin{aligned}
&H_0^{(N)}\psd \,, & H_1^{(N)}\psd \,,\\
&H_0^{(N-1)}-(x_0+R)H_1^{(N)}\psd\,, \qquad & H_1^{(N-1)}-(x_0+R)H_2^{(N)}\psd \,.
\end{aligned}
\end{equation}
Here $H_M^{(N)}$ denotes the matrix
\begin{equation}\label{eq:Hankels}
H_M^{(N)}=
\left\{\int_R^\infty\frac{\dd\mu(t)}{(x_0+t)^{1+i+j+M}}\right\}_{i,j=0}^{\floor{(N-M)/2}}
=\left\{(-1)^{i+j+M}\frac{f^{(i+j+M)}(x_0)}{(i+j+M)!}\right\}_{i,j=0}^{\floor{(N-M)/2}}\,.
\end{equation}
Intuitively, the two inequalities in the first row of eq.~\eqref{eq:fourHankelineq} express that $\mu\geq0$, while those in the second row express the fact that $\mu$ is supported on $[R,\infty)$.

We now demonstrate these new inequalities on the toy example introduced above: $f(x) = \log\frac{x+2}{x+1}$ at $x_0 = 0$. Since the corresponding $
\mu$ is supported on $[1,\infty)$ and
$x_0=0$, the shift is $x_0+R=1$ and the entries of $H_0^{(N-1)} - (x_0+R)H_1^{(N)}$ are
$a_k-a_{k+1}=\int_1^2 t^{-(k+2)}(t-1)\,\dd t\geq0$. At $N=3$ this becomes
\[
\begin{pmatrix}a_0-a_1&a_1-a_2\\a_1-a_2&a_2-a_3\end{pmatrix}\psd
\quad\Longleftrightarrow\quad
f(x_0) \;\geq\;a_1+\frac{(a_1-a_2)^2}{a_2-a_3}=\frac{11}{16} \,,
\]
which is already tighter than the $2/3$ obtained from $H^{(2)}_0$.

As $N$ increases, this bound becomes increasingly tight. In fact, for a univariate function as considered here, it is well established that the four linear matrix inequalities \eqref{eq:fourHankelineq} suffice to cut out the truncated moment cone. In particular, in the variable $s = 1/(x_0 + t) \in \bigl(0, 1/(x_0 + R)\bigr]$ this is equivalent to the classical Hausdorff moment problem, which is determinate because the support is bounded \cite{Widder:1941,Curto:1991rec}. The following theorem makes this precise.

\begin{tcolorbox}
\begin{theorem}[Hankel characterization of Stieltjes functions]
For a function $f(z)$ analytic on  $\mathbb{C}\setminus (-\infty, -R]$, the following are equivalent.
\begin{itemize}
  \item $f(z)$ is a Stieltjes function with corresponding positive measure $\mu(t)$ supported on $[R,\infty)$.
  \item For every real $x_0>-R$, the Taylor coefficient sequence $a_k=(-1)^k\frac{f^{(k)}(x_0)}{k!}$ satisfies the inequalities \eqref{eq:fourHankelineq} for all $N$.
\end{itemize}
\end{theorem}
\end{tcolorbox}

At any fixed $N$ it suffices to consider only two of the four inequalities. In particular, for even $N$ we have
\[
H_1^{(N)} = \left(H_1^{(N-1)} - (x_0 + R)H_2^{(N)} \right) + (x_0 + R)(H_0^{(N)})_{\{2,\ldots, \floor{N/2}+1 \} }\,,
\]
where $(H_0^{(N)})_{\{2,\ldots, \floor{N/2} \} }$ denotes the principal submatrix of $H_0^{(N)}$ obtained by deleting the first row and column. It follows that for even $N$
the inequalities
\begin{equation}
H_0^{(N)} \psd \,, \qquad H_1^{(N-1)}-(x_0+R)H_2^{(N)}\psd 
\end{equation}
imply all four inequalities in \eqref{eq:fourHankelineq}, while for odd $N$ the remaining two inequalities
\begin{equation}
H_1^{(N)}\psd \,, \qquad H_0^{(N-1)}-(x_0+R)H_1^{(N)}\psd
\end{equation}
suffice to cut out the truncated moment cone by an analogous argument. Thus in practice we only need to use two of the four inequalities when considering any fixed $N$.

The same constraints apply to generalized Stieltjes functions. If 
\[
f(x)=\int_R^\infty (x+t)^{-\lambda}\,\dd\mu(t)
\]
with $\lambda>0$, as for the integrals of section~\ref{sec:construction}, then
\begin{equation}
  a_k=\frac{(-1)^k}{k!}f^{(k)}(x_0)=\frac{\Gamma(\lambda+k)}{\Gamma(\lambda)~k!}\int_R^\infty\frac{\dd\mu(t)}{(x_0+t)^{\lambda+k}}\,,
\end{equation}
so the rescaled coefficients $\tilde a_k=k!\,a_k\, \Gamma(\lambda)/\Gamma(\lambda+k)$ are the sign-stripped Taylor coefficients at \(x_0\) of the ordinary Stieltjes function with the positive measure $(x_0+t)^{1-\lambda}\dd\mu(t)$, and the inequalities~\eqref{eq:fourHankelineq} hold for the sequence $\tilde a_k$. This is what we impose throughout, with $\lambda=1+\eps$ for the box of section~\ref{sec:massive-box} and $\lambda=1+3\eps$ for the three-loop family of section~\ref{sec:examples}. For $\lambda<1$ it is stronger than imposing the inequalities on $a_k$ alone, which also hold since a generalized Stieltjes function of order $\lambda<1$ is an ordinary Stieltjes function.

The rescaled constraints also apply to generalized Stieltjes functions with $\lambda >1$. This could be useful, since it offers additional freedom in choosing the master integral basis. We leave this to future work.

\subsection{From differential equations to semidefinite programming}
\label{sec:diff-eq-sdp}
We now show how to apply the positivity properties of Stieltjes functions to simultaneously bound a vector of Stieltjes Feynman integrals which are related by a first-order differential-equation system \cite{Kotikov:1990kg,Bern:1993kr,Remiddi:1997ny,Gehrmann:1999as} with rational coefficients.

Let $\vect g=(g_1,\ldots,g_n)$ be a vector of master integrals depending on the kinematic variables $u_1,\ldots,u_p$ and satisfying the first-order differential-equation system
\begin{equation}
\label{eq:gen_diff_eq}
  \partial_u\vect g=A_u \, \vect g \,,
\end{equation}
with rational matrices $A_u$.
Suppose that each master integral $g_j$ is a Stieltjes function of each variable $u$ separately. Furthermore, let $x_0=(u_1^0,\ldots,u_p^0)$ be a base point at which one component is known, say $g_n(x_0)=1$. 
Our goal is to approximate the remaining values $g_1(x_0),\ldots,g_{n-1}(x_0)$, which we collect into the vector of unknowns $\vect g(x_0)\in\mathbb{R}^{n-1}$.

Since each master integral $g_\ell$ is Stieltjes in each variable $u$, the sequence of Taylor coefficients of the expansion of $g_\ell$ with respect to each kinematic variable $u$, namely
\[
a_{u,k}^{(\ell)}=\frac{(-1)^k}{k!}\partial_u^kg_\ell(x_0) \,,
\]
must satisfy the four types of inequalities~\eqref{eq:fourHankelineq}. Since $\vect g$ satisfies the differential equation~\eqref{eq:gen_diff_eq}, we have that $\partial_u^k\vect g=A_u^{(k)}\vect g$, where $A_u^{(k)}$ is given by the recursion
\begin{equation}\label{eq:recursionA}
A_u^{(1)}=A_u \,,\qquad A_u^{(k)}=A_u^{(k-1)}A_u+\partial_uA_u^{(k-1)} \,.
\end{equation}
We thus see that 
\[
a_{u,k}^{(\ell)}=\frac{(-1)^k}{k!}\bigl(A_u^{(k)}(x_0)\vect g(x_0)\bigr)_\ell
\]
are linear functions of the unknowns $\vect g(x_0)$ with rational coefficients. Moreover, building $A_u^{(k)}$ requires only $A_u$ itself. In practice, every variable other than $u$ may be set to its numerical value before any derivative is taken, so only single-variable differentiation of a rational function is needed.  
We denote by $S_N$ the set of all such values of $\vect g(x_0)$ which satisfy the PSD constraints of \eqref{eq:fourHankelineq} at truncation order $N$. This set forms a \emph{spectrahedron}, meaning a subset of $\mathbb{R}^{n-1}$ which is cut out by linear matrix inequalities.

We again use a simple example to illustrate this procedure. Consider the function
\begin{equation}
f(u,v)=\int_0^\infty\frac{\dd t}{(u+t)(v+t)}=\frac{\log(u/v)}{u-v} \,,
\end{equation}
which is Stieltjes in $u$ and $v$ separately. In particular, it has measures $\dd t/(v+t)$ and $\dd t/(u+t)$ respectively, which are both positive on $[0,\infty)$. Thus the branch point is $R=0$ in both variables. In the following, we consider the base point $(u_0,v_0)=(1,2)\in(0,\infty)\times(0,\infty)$ and write $y=f(u_0,v_0)$ for the single unknown we wish to approximate.

The function satisfies the differential-equation system
\begin{equation}
\partial_u\begin{pmatrix}f\\1\end{pmatrix}
=\begin{pmatrix}-\frac1{u-v}&\frac1{u(u-v)}\\0&0\end{pmatrix}
\begin{pmatrix}f\\1\end{pmatrix} \,,\qquad
\partial_v\begin{pmatrix}f\\1\end{pmatrix}
=\begin{pmatrix}\frac1{u-v}&-\frac1{v(u-v)}\\0&0\end{pmatrix}
\begin{pmatrix}f\\1\end{pmatrix} \,.
\end{equation}
We first compute the coefficients $a_{v,k}$ with respect to variable $v$. Following the discussion above, we substitute $u=u_0=1$ into $A_v$ before differentiating, so that only one variable remains symbolic:
\begin{equation}
A_v=\begin{pmatrix}\frac1{1-v}&\frac1{v(v-1)}\\0&0\end{pmatrix}\,.
\end{equation}
The recursion~\eqref{eq:recursionA} then gives
\begin{equation}
A_v^{(1)}=A_v \,,\qquad
A_v^{(2)}=\begin{pmatrix}\frac2{(v-1)^2}&\frac{1-3v}{v^2(v-1)^2}\\0&0\end{pmatrix}\,,\qquad
A_v^{(3)}=\begin{pmatrix}\frac{-6}{(v-1)^3}&\frac{2-7v+11v^2}{v^3(v-1)^3}\\0&0\end{pmatrix},\ \cdots
\end{equation}
Note that the second row of every $A_v^{(k)}$ vanishes since the second component is constant. Evaluating these matrices at $v=v_0=2$, applying these to the vector $\vect g(x_0)=(y,1)^T$ and computing $a_{v,k}$ yields
\begin{equation}
a_{v,0}=y\,,\quad a_{v,1}=y-\tfrac12 \,,\quad
a_{v,2}=y-\tfrac58 \,,\quad
a_{v,3}=y-\tfrac23 \ \cdots
\end{equation}
Repeating this procedure for $u$, where now $v=v_0=2$ is substituted first, gives
\begin{equation}
a_{u,0}=y \,,\qquad a_{u,1}=1-y\,,\qquad a_{u,2}=y-\tfrac12 \,,\qquad a_{u,3}=\tfrac56-y \, \cdots
\end{equation}
Thus we see that every Taylor coefficient is a linear function of the single unknown $y=f(1,2)$. Observe also that these coefficients were obtained without ever evaluating $f$.

The target numerical point is $y=f(1,2)=\log2\approx0.693147$. Now we use the inequalities \eqref{eq:fourHankelineq} to approximate this value. Since $R=0$ and $v_0=2$, we have $v_0+R=2$, and the first inequality in \eqref{eq:fourHankelineq} gives
\begin{equation}
H_0^{(2)}(1,v)=\begin{pmatrix}y&y-\tfrac12\\y-\tfrac12&y-\tfrac58\end{pmatrix}\psd
\quad\Longleftrightarrow\quad y\geq\frac23 \,.
\end{equation}
On the other hand, the localizing constraint $H_1^{(1)}-2H_2^{(2)}\psd$ yields
\begin{equation}
H_1^{(1)}(1,v)-2H_2^{(2)}(1,v)=a_{v,1}-2a_{v,2} \geq 0
\quad\Longleftrightarrow\quad y\leq\frac34 \,.
\end{equation}
Hence $S_2=[\tfrac23,\tfrac34]$, which indeed contains $\log2$. Increasing $N$ tightens this rapidly:
\begin{equation}
S_4=\left[\frac9{13},\frac{25}{36}\right],\qquad
S_6=\left[\frac{131}{189},\frac{61}{88}\right],\qquad S_8=\left[\frac{445}{642},\frac{3743}{5400}\right].
\end{equation}
The intervals $S_2$, $S_4$, $S_6$ and $S_8$ have widths of approximately $8.3\cdot10^{-2}$, $2.1\cdot10^{-3}$, $6.0\cdot10^{-5}$ and $1.7\cdot 10^{-6}$, respectively. As $N$ increases, this yields a descending sequence of intervals which give increasingly tight bounds on $f(1,2)$. Observe that in this case $S_2$ is already bounded and $S_N \subseteq S_{N'}$ for any $N \geq N'$.

At this point, it is worth making a brief comparison with the constraints obtained from the CM bootstrap of ref.~\cite{Ditsch:2025dhp}. A function $h$ on $(-R,\infty)$ is called CM if
$(-1)^n h^{(n)}(x)\geq0$ for all $n\in\mathbb{N}_0$ and $x>-R$. By the
Bernstein--Widder theorem~\cite{Widder:1941}, this is equivalent to
\begin{equation}
  h(x)=\int_0^\infty e^{-(x+R)s}\,\dd\nu(s)\,,
  \qquad \dd\nu(s)\geq0\,.
  \label{eq:bernstein_widder_shifted}
\end{equation}
Every Stieltjes function is CM on $(-R,\infty)$, since
\begin{equation}
  (-1)^n f^{(n)}(x)
  =n!\int_R^\infty\frac{\dd\mu(t)}{(x+t)^{n+1}}\geq0\,.
  \label{eq:stieltjes_cm}
\end{equation}
The converse is not true. At fixed $x>-R$, define
\begin{equation}
  d_n(x)=(-1)^n h^{(n)}(x)\,,
  \qquad
  K_M^{(N)}(x)
  =\bigl(d_{i+j+M}(x)\bigr)_{i,j=0}^{\lfloor(N-M)/2\rfloor}\,.
  \label{eq:cm_hankel_definition}
\end{equation}
The CM representation implies
\begin{equation}
  K_0^{(N)}(x)\succeq0\quad(N\geq0)\,,
  \qquad
  K_1^{(N)}(x)\succeq0\quad(N\geq1)\,.
  \label{eq:cm_hankel}
\end{equation}
There are no CM analogues of the last two inequalities in
eq.~\eqref{eq:fourHankelineq}, because the support of $\nu$ in the Laplace variable $s \in [0, \infty)$ is unbounded. At second order, the CM property of $h(x)$ and Stieltjes property of $f(x)$ imply, respectively,
\begin{align}
  h(x)h''(x)-h'(x)^2&\geq0\,,\label{eq:cm_second_order}
 \\
  \frac{1}{2}f(x)f''(x)-f'(x)^2&\geq0\,.
  \label{eq:stieltjes_second_order}
\end{align}
The latter is stronger and is accompanied by the two additional matrix
inequalities. These provide information beyond a bootstrap based only on
complete monotonicity~\cite{Ditsch:2025dhp}.

Returning to our example, by eq.~\eqref{eq:cm_hankel} and in the direction $v$, the leading condition is
\begin{equation}
K_0^{(2)}=\begin{pmatrix} y & y-\tfrac12\\[2pt] y-\tfrac12 & 2y-\tfrac54\end{pmatrix}
\succeq0
\quad\Longleftrightarrow\quad
y^2-\tfrac14y-\tfrac14\geq0
\quad\Longleftrightarrow\quad
y\geq\tfrac18\bigl(1+\sqrt{17}\bigr) \,,
\end{equation}
while in the direction $u$ the first derivative alone gives $d_{u,1}=1-y\geq0$.
Complete monotonicity therefore does bound $\log2$ from both sides, but only as
\begin{equation}
0.6403\ldots=\tfrac18\bigl(1+\sqrt{17}\bigr)\;\leq\;\log2\;\leq\;1 ,
\end{equation}
a width of $0.36$ against the $8.3\cdot10^{-2}$ of $S_2=[\tfrac23,\tfrac34]$. The gap continues to grow with $N$. In particular at $N=4,6,8$ the complete-monotonicity widths are $3.1\cdot10^{-2}$, $3.0\cdot10^{-3}$ and $3.0\cdot10^{-4}$, against $2.1\cdot10^{-3}$, $6.0\cdot10^{-5}$ and $1.7\cdot10^{-6}$ above, so by $N=8$
the two enclosures differ by more than two orders of magnitude.

For a general differential-equation system, each constraint in \eqref{eq:fourHankelineq} is a linear matrix inequality in $\vect g(x_0)$. Bounding a single value $g_\ell(x_0)$ thus amounts to optimizing a linear functional over $S_N(x_0)$, that is, to solving the semidefinite programs (SDPs)
\begin{equation}
\max_{\vect g(x_0)\in S_N(x_0)}g_\ell(x_0)
\qquad\text{and}\qquad
\min_{\vect g(x_0)\in S_N(x_0)}g_\ell(x_0) \,.
\end{equation}
Solving both yields an interval which is guaranteed to contain $g_\ell(x_0)$, and these intervals shrink as $N$ increases. Observe that the constraints defining $S_N$ do not depend on which $\ell$ we bound, so they need only be assembled once and may then be reused for all $2(n-1)$ optimizations. The two SDPs above can be solved in polynomial time with an interior-point method. Each SDP has $n-1$ decision variables, namely the function values to be bounded. The constraints come in $2np$ blocks, one for each combination of the two inequality types selected by the parity of $N$, the $n$ functions $g_\ell(x)$, and the variables $u_1,\ldots,u_p$ of the differential equation. Each block is a square matrix with at most $\floor{N/2}+1$ rows and columns.

We emphasize that the differential equations need not be written in a Stieltjes basis. Let $\vect I$ be any basis of the family, for instance one produced by the Laporta algorithm, with $\partial_u\vect I=A_u\vect I$, where $A_u$ now denotes the connection matrix of the basis $\vect I$, and let $\vect h=B\,\vect I$ be a list of $m\geq n$ Stieltjes integrals expressed in that basis by IBP reduction, with $B$ an $m\times n$ matrix of rational functions. The derivatives of $\vect h$ follow from
\begin{equation}
  C_{u,0}=B\,,\qquad
  C_{u,k+1}=\partial_u C_{u,k}+C_{u,k}A_u\,,\qquad
  \partial_u^k\vect h=C_{u,k}\,\vect I\,,
  \label{eq:laporta-candidate-derivatives}
\end{equation}
so that the Taylor coefficients $a^{(\ell)}_{u,k}=\frac{(-1)^k}{k!}[C_{u,k}(x_0)\vect I(x_0)]_\ell$ of every $h_\ell$ are linear in the unknown values $\vect I(x_0)$, exactly as before. Positivity is imposed on the elements of \(\vect h\), whereas the elements of
\(\vect I\) need not themselves be Stieltjes. Taking a list longer than the basis can strengthen the
constraints without introducing additional decision variables, since the
unknown vector remains \(\vect I(x_0)\in\mathbb{R}^n\). It does, however, increase the cost of
solving the resulting SDPs. There is consequently a trade-off between the
strength of the positivity constraints and the size of the semidefinite
program.

We collect the preceding discussion into the {\it Algorithm: Stieltjes bootstrap}. Before discussing the complexity of our algorithm and its implementation, we first demonstrate it on a concrete physical example.

\begin{tcolorbox}[float, floatplacement=tbp,title=Algorithm:\ Stieltjes bootstrap]
\SetKwInOut{Input}{Input}\SetKwInOut{Output}{Output}
\Input{A first-order linear system $\partial_u\vect g=A_u\vect g$ in variables $u_1,\ldots,u_p$ with rational coefficients, such that each $g_\ell$ is Stieltjes in each $u$; a base point $x_0$ in the Euclidean region; branch values $R_{\ell,u}$; a normalization $g_n(x_0)=1$; a target accuracy of $d_{\rm t}$ digits; a maximum order $N_{\max}$.}
\Output{Intervals $[\underline g_\ell,\overline g_\ell]\ni g_\ell(x_0)$ for $1\leq\ell\leq n-1$ whose endpoints agree to at least $d_{\rm t}$ digits, or a report that $N_{\max}$ was reached.}
$N\leftarrow2$\;
\While{$N\leq N_{\max}$}{
  \ForEach{$u\in\{u_1,\ldots,u_p\}$}{
    substitute every variable other than $u$ at its value in $x_0$\;
    $A_u^{(1)}\leftarrow A_u$\;
    \For{$k\leftarrow2$ \KwTo $N$}{
      $A_u^{(k)}\leftarrow A_u^{(k-1)}A_u+\partial_uA_u^{(k-1)}$\;
    }
    evaluate $A_u^{(k)}$ at $x_0$ for $1\leq k\leq N$\;
    \ForEach{$1\leq\ell\leq n$}{
      $a_{u,0}^{(\ell)}\leftarrow(\vect g(x_0))_\ell$\;
      $a_{u,k}^{(\ell)}\leftarrow\dfrac{(-1)^k}{k!}\bigl(A_u^{(k)}(x_0)\vect g(x_0)\bigr)_\ell$ for $1\leq k\leq N$\;
    }
  }
  Obtain $S_N(x_0)$ by assembling the two relevant inequalities of the four families of eq.~\eqref{eq:fourHankelineq} based on the parity of $N$, with thresholds $R_{\ell,u}$, for every $\ell$ and $u$\;
  \ForEach{$1\leq\ell\leq n-1$}{
    $\underline g_\ell\leftarrow\min\{g_\ell(x_0):\vect g(x_0)\in S_N(x_0)\}$\;
    $\overline g_\ell\leftarrow\max\{g_\ell(x_0):\vect g(x_0)\in S_N(x_0)\}$\;
  }
  \If{$\min_\ell\operatorname{agree}(\underline g_\ell,\overline g_\ell)\geq d_{\rm t}$}{
    \Return{$\{[\underline g_\ell,\overline g_\ell]\}_{\ell=1}^{n-1}$}\;
  }
  $N\leftarrow N+2$\;
}
\Return{the current intervals, together with a warning that $N_{\max}$ was reached}\;
\end{tcolorbox}

\subsection{A pedagogical example: the one-loop massive box}
\label{sec:massive-box}

We now apply the Stieltjes bootstrap to the one-loop box family which was studied analytically in \cite{CaronHuot:2014lda}. This integral family takes the form
\begin{equation}\label{eq:box_def}
I_{a_1a_2a_3a_4}^{(D)}(s,t,m^2)=\int\frac{\dd^Dk}{\ii\pi^{D/2}}\frac1{D_1^{a_1}D_2^{a_2}D_3^{a_3}D_4^{a_4}} \,,
\end{equation}
with
\begin{equation}
\begin{aligned}
D_1&=-k^2+m^2 \,,&D_2&=-(k-p_1)^2+m^2 \,, \\
D_3&=-(k-p_1-p_2)^2+m^2 \,, \quad &D_4&=-(k+p_4)^2+m^2 \,.
\end{aligned}
\end{equation}
The superscript in $I_{a_1a_2a_3a_4}^{(D)}(s,t,m^2)$ records the spacetime dimension. To apply our main algorithm, we must find a Stieltjes basis, that is a set of master integrals all of which are Stieltjes. We first enumerate the admissible Stieltjes integrals. As explained in section~\ref{sec:construction}, we look for scalar integrals $I_{a_1a_2a_3a_4}^{(D_0-2\eps)}$, with even $D_0$, whose Stieltjes exponent is $\lambda=1+\eps$. At $\eps=0$, this means
\begin{equation}
\sum_{e=1}^4 a_e-\frac{D_0}{2}=1.
\end{equation}
We also require ultraviolet finiteness, i.e.
\begin{equation}
\sum_{e\in\gamma}a_e-\frac{L_\gamma D_0}{2}>0
\end{equation}
for the full graph and for every subgraph $\gamma$. These linear conditions on the $a_e$ define a small integer linear program which can be quickly solved. The resulting higher-dimensional integrals are then mapped back to the standard $D=4-2\eps$ basis using the dimensional recurrence relations of ref.~\cite{Tarasov:1996br}, as implemented for example in LiteRed \cite{Lee:2013mka}. In practice, passing from a conventional Laporta basis to a Stieltjes basis amounts to one further linear change of basis, whose cost is negligible compared with the derivation of the differential equations. Solving the integer program and applying dimensional recurrence, we obtain the Stieltjes basis
\begin{equation}
\vect g=\bigl(I_{1111}^{(6)},I_{1110}^{(4)},I_{1101}^{(4)},I_{1010}^{(2)},I_{0101}^{(2)},I_{0003}^{(4)}\bigr) \,.
\end{equation}
These master integrals satisfy a system of differential equations,
\begin{equation}
\frac{\partial\vect g}{\partial m^2}=A_{m^2} \, \vect g\, ,
\qquad \frac{\partial\vect g}{\partial s}=A_s \, \vect g\, ,
\qquad \frac{\partial\vect g}{\partial t}=A_t \, \vect g\, .
\end{equation}
The connection matrices $A_{m^2},A_s,A_t$ are sparse $6\times6$ matrices whose entries are rational functions of $s,t,m^2,\eps$. 
They are given in appendix~\ref{app:matrices}.

\begin{figure}[!t]
\centering
\includegraphics[width=0.62\textwidth]{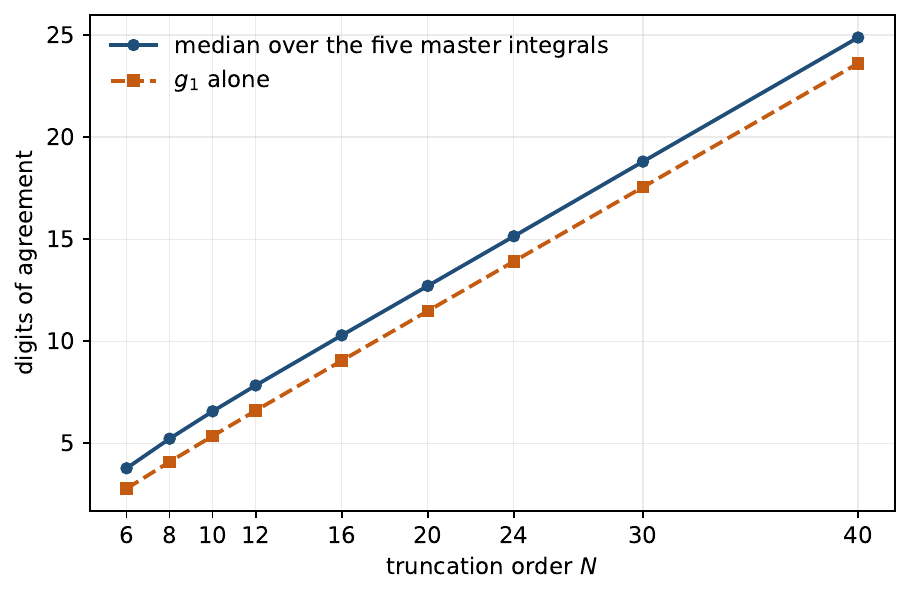}
\caption{The width of the enclosure for the one-loop box against the truncation order $N$, on a logarithmic scale, for the median over the five bounded master integrals and for $g_1$ alone. The width decays to zero exponentially rather than polynomially.}
\label{fig:box_convergence}
\end{figure}

We now apply the Stieltjes-bootstrap algorithm at the Euclidean point
\begin{equation}
    (s_0,t_0,m^2)=(-3,-7,1) \,.
\end{equation}
We normalize the sixth master integral so that $g_6(x_0)=1$, leaving five unknown master-integral values.  At each truncation order, the necessary Hankel inequalities given in eq.~\eqref{eq:fourHankelineq} based on the parity of $N$ are assembled for every master integral and for each of the three Stieltjes variables $s$, $t$, and $m^2$. Table~\ref{tab:1} records the number of digits of agreement between the upper and lower bound of each bounding interval,
\[
  \mathrm{agree}(\underline g, \overline g) \;=\; -\log_{10}\frac{\overline g-\underline g}{\tfrac{1}{2}\lvert \underline g+\overline g\rvert} \,.
\]
For each master, including $g_1$, the bootstrap provides intervals which contain the known analytic value and whose widths decay exponentially to zero, see figure~\ref{fig:box_convergence}. The median accuracy over the five unknown master integrals grows by approximately
$0.6$ digits per unit increase in $N$, reaching more than $24$ digits at $N=40$. 

The Stieltjes property also permits a complementary evaluation, based on the soft expansion and Padé approximants~\cite{Ditsch:2025dhp}, which we use as a check.
For the one-loop box in six dimensions, denoted by $I$, we introduce a one-dimensional path in kinematic space as
\begin{equation}
    I(x) \coloneq I(xs,xt,m^2) \,,
    \qquad
    I(0)=\frac{1}{6m^2} \,,
    \qquad
    I(1)=I(s,t,m^2) \,.
\end{equation}
The point $x=0$ corresponds to a soft limit, in which the box reduces to a tadpole, and the expansion near it is
\begin{equation}
    I(x)=\frac{1}{m^2}\sum_{k=0}^{\infty}~(-1)^k~c_k \, x^k \,,
\end{equation}
where
\begin{equation}
  c_k = \sum_{j=0}^k \frac{\Gamma(1+k)\,\Gamma(1+k-j)\,\Gamma(1+j)}{\Gamma(2k+4)}\,
  \left(\frac{-s}{m^2}\right)^j \left(\frac{-t}{m^2}\right)^{k-j}\,.
  \label{eq:ak_box}
\end{equation}
The coefficients $c_k$ then determine the Padé approximants $\mathcal{P}_N^{\pm}(x)$
introduced in appendix~\ref{sec:stieltjes_review}, whose Stieltjes bounds satisfy
\begin{equation}
    \mathcal{P}_N^-(x)\leq I(x)\leq \mathcal{P}_N^+(x)\,,
    \qquad x>0 \,.
\end{equation}
At the point $(s,t,m^2) = (-3,-7,1)$ the $[10/10]$ and $[9/10]$ Pad\'e approximants constructed from twenty-one
Taylor coefficients reproduce $g_1$ to 12.5 digits, to be compared with the 11.5 digits obtained
from the SDP at $N = 20$ in Table~\ref{tab:1}.

Compared with the SDP method, the soft-expansion--Padé method starts from explicitly known moments at a simple boundary point, and transports this information
along the path from $x=0$ to $x=1$.
It approximates one integral at a time from explicit boundary data,
whereas the SDP simultaneously produces certified
enclosures for the full vector of master integrals without assuming
their boundary values, apart from a single normalization.  The two methods are therefore complementary: when a sufficiently long soft expansion is readily available, Padé
provides an inexpensive way of propagating it over the Euclidean region; when such boundary information is unavailable or difficult
to compute, the SDP can instead determine the required local data from the differential equations and positivity.  In a combined
strategy, SDP bounds may supply boundary values, while Padé approximants subsequently transport them to other kinematic points.

For a general family the soft expansion may be generated by the differential equation itself: the Taylor coefficients of the master integrals at the soft point follow recursively from the connection matrix and the
boundary values, so that only the boundary values have to be computed, and the Stieltjes elements are obtained
from the Laporta basis by the reduction matrix.
We use this in Table~\ref{tab:lad2-comparison} to evaluate the three-loop family of section~\ref{sec:examples} from its soft limit.

\begin{table}[!t]
\centering
\begin{tabular}{rrrrc}
\toprule
$N$ & block size & median digits & digits on $g_1$ & seconds\\
\midrule
 6 &  4 &  3.77 &  2.77 &   1.1 \\
 8 &  5 &  5.22 &  4.08 &   1.6 \\
10 &  6 &  6.56 &  5.35 &   2.2 \\
12 &  7 &  7.83 &  6.59 &   3.0 \\
16 &  9 & 10.28 &  9.04 &   5.1 \\
20 & 11 & 12.71 & 11.48 &   8.0 \\
24 & 13 & 15.14 & 13.91 &  11.7 \\
30 & 16 & 18.80 & 17.55 &  19.6 \\
40 & 21 & 24.88 & 23.61 &  38.1 \\
\bottomrule
\end{tabular}
\caption{Convergence of the Stieltjes bootstrap for the one-loop box at $s=-3$, $t=-7$, $m^2=1$, with threshold $R=4$ in $-s$ and $-t$, run in arbitrary precision. The median of the agreement between the lower and upper bound is taken over the five functions and grows by roughly 0.6 digits per unit of $N$. Note that this timing refers only to the time it takes to solve the SDPs once the blocks are built.}
\label{tab:1}
\end{table}

\subsection{Discussion of convergence properties}
\label{sec:convergence}
In the examples of Sections~\ref{sec:diff-eq-sdp} and~\ref{sec:massive-box}, the width of the bounds decays exponentially with $N$. This is not always the case. The differential equation determines the master integrals only up to homogeneous solutions, and the bounds can only converge to the true values when positivity excludes all of them. In the examples we studied, we observe three types of behavior. First, if a homogeneous solution is itself Stieltjes, it cannot be excluded, and the upper and lower bounds do not converge to one another. Second, in the examples where no homogeneous solution is Stieltjes but at least one is singular only on the Stieltjes branch cut $(-\infty,-R]$, we observe algebraic convergence. Third, in the examples where every homogeneous solution has a singularity away from the cut, we observe exponential convergence.

The massless one-loop box with on-shell legs illustrates the first two cases. In $D=6$, with $x=t/s$, it is proportional to
\begin{equation}
  f(x)=\frac{\ln^2 x+\pi^2}{1+x}\,,
  \label{eq:massless-box-D6}
\end{equation}
which satisfies
\begin{equation}
  \partial_x\begin{pmatrix}f\\ \ln x\\ 1\end{pmatrix}
  =\begin{pmatrix}-\frac{1}{1+x}&\frac{2}{x(1+x)}&0\\[1mm]0&0&\frac1x\\[1mm]0&0&0\end{pmatrix}
  \begin{pmatrix}f\\ \ln x\\ 1\end{pmatrix}\,.
  \label{eq:massless-box-D6-de}
\end{equation}
With $\ln x$ supplied as input, the only homogeneous solution is $1/(1+x)$, which is itself Stieltjes, so that the bootstrap does not determine a unique solution. In $D=6-2\eps$ with $\eps<0$, the diagonal entry $-1/(1+x)$ becomes $-\eps/x-(1-\eps)/(1+x)$, and the homogeneous solution $x^{-\eps}(1+x)^{\eps-1}$ is too singular at $x=-1$ to be Stieltjes of order $\lambda=1+\eps$. The bootstrap then determines the box uniquely, and we observe algebraic convergence. In this example, the limits $N\to\infty$ and $\eps\to0$ do not commute.

The massive box of section~\ref{sec:massive-box} illustrates the third case. Its homogeneous solutions are singular at points in the Euclidean region, i.e.\ off the cut, and are therefore excluded. The width of the bounds then shrinks exponentially.

We leave a more detailed analysis of the convergence properties of the Stieltjes bootstrap, including whether this pattern holds in general, to future work.

\subsection{Complexity and implementation}
\label{sec:complexity}

We have implemented the Stieltjes bootstrap algorithm in both \texttt{Python} and \texttt{Mathematica}. The first main component of this algorithm is the construction of the $2np$ blocks which correspond to the two of the four PSD constraints in eq.~\eqref{eq:fourHankelineq} that are needed based on the parity of $N$, the $n$ functions to optimize, and the $p$ connection matrices encoding the derivative with respect to $u_1, \ldots, u_p$. As discussed in section~\ref{sec:diff-eq-sdp}, when computing the blocks corresponding to $u_i$, we evaluate every other $u_j$ to its fixed value. This means that $A_{u_i}$ becomes a matrix whose entries are univariate rational functions which helps keep the expressions small. In all of the examples we have computed, this process does not require much memory, even when the input connection matrices are quite large and dense. Our \texttt{Python} implementation of this build step is fully parallelized and can easily be deployed on a larger cluster.

The second main component is the maximization and minimization of each coordinate over the spectrahedron $S_N$. Our implementation has several available solvers for this which include \texttt{Clarabel} \cite{Goulart2026Clarabel}, \texttt{MOSEK} \cite{mosek}, \texttt{sdpa\_gmp} \cite{Nakata2010SDPAGMP}, and \texttt{sdpb} \cite{Landry:2019qug, Simmons-Duffin:2015qma}. The first two are called directly through the \texttt{Python} package \texttt{cvxpy} \cite{diamond2016cvxpy} for convex programming. Both of these solvers are limited to double precision, whereas \texttt{sdpa\_gmp} and \texttt{sdpb} are capable of arbitrary precision. In most examples, we have found that \texttt{Clarabel} and \texttt{MOSEK} plateau at around five to six digits of agreement between the max and min. This is most likely because the blocks of our semidefinite program are formed from Hankel matrices whose condition number grows exponentially in their size. Obtaining tight bounds, i.e.\ intervals for each $g_\ell(x_0)$ that agree to at least ten digits, therefore requires solving the SDPs in very high precision, and we primarily use \texttt{sdpa\_gmp} and \texttt{sdpb} for the physical examples discussed in this paper. Just like the build step, this phase parallelizes well. If $\vect g = (g_1, \ldots, g_n)$, then we must solve $2(n-1)$ SDPs, since for each of the functions that are not normalized we must compute both its maximum and minimum over $S_N$; our implementation runs them in parallel. Thus the wall-clock time of the Stieltjes bootstrap is usually determined by the cost of solving a single SDP.

All of the solvers that our implementation utilizes rely on \emph{primal--dual interior-point
methods} to solve the SDPs. These methods keep their iterates strictly inside the cone of
positive definite block-diagonal matrices and follow the central path of the log-determinant
barrier towards the optimum by taking Newton steps organized as predictor--corrector pairs. In our
case we must solve $2(n-1)$ SDPs whose objectives are simply the coordinate functions. As described in the previous sections, each pair of a function and direction contributes two blocks, of sizes $\lfloor N/2\rfloor+1$ and $\lceil N/2\rceil$. Thus the cost of a single interior-point step is
\begin{align}
  \mathcal O\Bigl((n-1)\sum_b s_b^3+(n-1)^2\sum_b s_b^2+(n-1)^3\Bigr)
 % =\mathcal O\bigl(n^2pN^3+n^3pN^2\bigr)
 =\mathcal O\bigl(n^2pN^2(N+n)\bigr),
\end{align}
where $s_b$ are the block sizes and the sums run over at most $2np$ blocks. Since all blocks have size about $N/2$, the first two terms exchange dominance at $N\approx2(n-1)$. These counts are arithmetic operations at the working precision, which in practice has to grow with $N$. The number of steps needed to converge to an optimum is governed by the total block size and the necessary precision. More precisely, to compute the endpoints to $\delta$ accuracy, in the worst case
$\mathcal O\bigl(\sqrt{npN}\,\log(npN/\delta)\bigr)$
steps are required, though in practice it is typically far fewer \cite{Boyd:2004}.

It is important to note that the moments $a_{u,k}^{(\ell)}$, and hence the constraints, are assembled once per $N$ and reused across all $2(n-1)$ optimizations. We also remark that when the connection matrices $A_u$ are sparse, both the build step and the solving of the SDPs become much faster. In particular, if a block involves only the last function $g_n$ which is normalized to 1, then it can be removed entirely. Similarly if a block involves only one $g_\ell$, i.e.\ it corresponds to a constraint of the form
\begin{align}
g_\ell M_1 + M_0 \psd \,,
\end{align}
such that $M_1$ is either positive or negative semidefinite, then that block may be replaced with $g_\ell \geq c$ or $g_\ell \leq c$, where $c$ is easily obtained by solving a generalized eigenvalue problem. When $A_u$ is sparse, such blocks naturally occur much more often and provide a significant computational payoff since this reduction can be performed once as a preprocessing step and thus every SDP becomes much simpler.

Lastly, we note that throughout this paper we often used \emph{certified digits}, which we now explain. Every enclosure we report is certified a posteriori in exact arithmetic. Besides approximate values of the integrals, the solver returns a dual solution which is a positive semidefinite combination of the constraints. This bounds the objective by weak duality. We do not rely on the solver's arithmetic
for this step. We convert its dual solution to exact rational numbers, verify
exactly that it is positive semidefinite, and recompute the bound it implies
exactly, rounding outward. Because of rounding, the solver's dual solution does not reproduce the
objective exactly. We account for the difference by assuming that every
unknown is at most $C$ in absolute value and subtracting the largest effect
the difference could then have on the objective. Thus the result is a valid bound
for every feasible point in this range regardless of the solver's accuracy. An inaccurate solve only weakens the bound. 
We then remove the assumption on $C$ by taking $C$ to be twice the largest
value absolute value of any of the returned coordinates and checking that every certified enclosure lies
strictly inside $[-C, C]$. Since the feasible set is convex and contains
points inside this range, a feasible point outside it would force another
feasible point onto the edge of the range, where the certified enclosures
rule it out. The bounds therefore hold without the assumption. The
certificate is exact for the problem data as passed to the solver.

%=====================================================================
\section{Applications to a three-loop two-mass Feynman integral family}
\label{sec:examples}
%=====================================================================
We now consider a three-loop self-energy family with two
internal masses. After defining the integrals and their
physical applications, we determine the kinematic domain
relevant to the positivity constraints. We then explain
how to formulate the bootstrap using a Laporta basis
together with a separate list of Stieltjes integrals,
and present the resulting bounds at a kinematic point relevant for the $Z$ and Higgs boson self-energies.

\subsection{Integrals and physical motivation}
\label{sec:lad2-setup}

Precision predictions for electroweak observables require
multi-loop self-energies with several internal masses.
These enter the relations between renormalized parameters
and measured quantities, including the gauge-boson masses
and the Fermi constant~\cite{Denner:2019vbn}.
In an on-shell scheme, mass and field counterterms involve
self-energies and their momentum derivatives. Their
evaluation therefore requires massive two-point integrals
both at zero momentum and at nonzero external invariants.

Recent progress includes the calculation of mixed
electroweak--QCD corrections of order
$\mathcal{O}(\alpha^2\alpha_s)$ to the prediction of the
$W$-boson mass from the Fermi constant measured in muon
decay~\cite{Dubovyk:2026nhx}. Here $\alpha$ and $\alpha_s$
are the electromagnetic and strong couplings, respectively.
This calculation requires three-loop self-energies at
nonzero external momentum. Purely electroweak
corrections at three-loop order will be required to match the experimental precision expected at proposed future electron-positron colliders such as CEPC, FCC-ee, and ILC~\cite{Chen:2020xzx,FCC:2025lpp}.

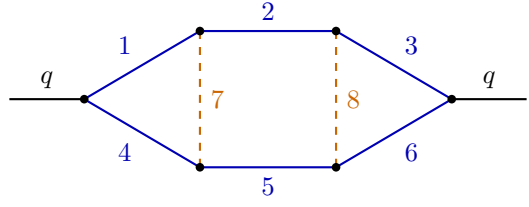
\begin{figure}[t]
\centering
\begin{tikzpicture}[
  scale=0.9,
  every node/.style={font=\small}
]
  \coordinate (L) at (-2.7,0);
  \coordinate (A) at (-1,1);
  \coordinate (B) at (1,1);
  \coordinate (C) at (-1,-1);
  \coordinate (D) at (1,-1);
  \coordinate (R) at (2.7,0);

  \draw[thick,blue!70!black]
    (L)--node[above left]{$1$}(A)
       --node[above]{$2$}(B)
       --node[above right]{$3$}(R);

  \draw[thick,blue!70!black]
    (L)--node[below left]{$4$}(C)
       --node[below]{$5$}(D)
       --node[below right]{$6$}(R);

  \draw[thick,dashed,orange!80!black]
    (A)--node[right]{$7$}(C);
  \draw[thick,dashed,orange!80!black]
    (B)--node[right]{$8$}(D);

  \draw[thick]
    (-3.8,0)--node[above]{$q$}(L);
  \draw[thick]
    (R)--node[above]{$q$}(3.8,0);

  \foreach \v in {L,A,B,C,D,R}
    \fill (\v) circle (1.8pt);
\end{tikzpicture}
\caption{The two-mass three-loop ladder. Solid blue lines
have mass $M_W$, and dashed orange lines have mass $m_r$.
The edge numbers correspond to the denominators in
eq.~\eqref{eq:lad2-propagators}. The irreducible scalar
product $D_9$ is not represented by an edge.}
\label{fig:lad2-family}
\end{figure}

Here we consider a family of three-loop integrals that contribute to neutral-boson self-energies in purely electroweak theory: the planar three-loop
ladder with six rail propagators of mass $M_W$ and two
rung propagators of mass $m_r$, shown in
figure~\ref{fig:lad2-family}. The choices $m_r=M_Z$ and
$m_r=M_H$ describe mass assignments occurring in bosonic
contributions to neutral-boson self-energies.
The scalar
integrals provide building blocks for these contributions
after tensor reduction. We retain the dependence on the
external momentum and the mass ratio, so that the same
family describes both mass assignments and their subsectors.

%Following the propagator-sign convention of
%section~\ref{sec:massive-box},
% JMH It's the standard sign convention :)
We define the scalar integrals of the family by
\begin{equation}
  I_{\vect a}^{(D)}
  =
  {\rm e}^{3\eps\gamma_E} (M_W^2)^{\sum_{j=1}^{9}a_j-3D/2}  \int\prod_{\ell=1}^{3}
  \frac{\dd^D k_\ell}{\ii \pi^{D/2}}\,
  \prod_{j=1}^{9}D_j^{-a_j}\,,
  \label{eq:lad2-family}
\end{equation}
with $\vect a=(a_1,\ldots,a_9)$ and
\begin{align}
\begin{aligned}
  D_i&=-k_i^2+M_W^2\,,
  &D_{i+3}&=-(q-k_i)^2+M_W^2\,, \\
  D_7&=-(k_1-k_2)^2+m_r^2\,, \qquad
  &D_8&=-(k_2-k_3)^2+m_r^2\,, \\
  D_9&=-(k_1-k_3)^2\,,
\end{aligned}
  \label{eq:lad2-propagators}
\end{align}
for $i=1,2,3$. The propagator denominators are understood as $D_j-\ii 0^+$
for $j=1,\ldots,8$, where $0^+$ is a positive infinitesimal. The irreducible scalar product $D_9$
occurs only in the numerator, i.e.\ $a_9\leq0$.
Vanishing propagator indices describe subsectors.
The prefactor $(M_W^2)^{\sum_j a_j-3D/2}$ in eq.~\eqref{eq:lad2-family} is conventional and removes the overall mass scale.
As a result, the integrals depend on two kinematic variables, which we define as
\begin{equation}
  x=\frac{q^2}{M_W^2}\,,\qquad \qquad
  z=\frac{m_r^2}{M_W^2}\,.
  \label{eq:lad2-dimensionless}
\end{equation}
Both choices of $m_r$ are described by the same IBP
reduction, which gives 60 master integrals in 35 sectors.
We performed the reduction independently with \textsc{FiniteFlow}~\cite{Peraro:2019svx} (with IBP relations generated by \textsc{LiteRed}~\cite{Lee:2013mka} and \textsc{NeatIBP}~\cite{Wu:2023upw}), with \textsc{FIRE}~\cite{Smirnov:2025prc} and with \textsc{Kira}~\cite{Lange:2025fba}.

We select scalar Stieltjes integrals as described in section~\ref{sec:construction}.
In $D=4-2\eps$ dimensions, the resulting integrals span only 17 of the 60 master integrals.
By contrast, working in $D=6-2\eps$ with $a_9=0$ and total propagator power ten yields 149 scalar Stieltjes integrals, from which we can construct a basis of 60 Stieltjes functions. Their Stieltjes exponent is $\lambda=1+3\eps$; therefore, requiring $0 < \lambda \leq 1$ restricts $\eps$ to $-1/3<\eps\leq0$. Dimensional recurrence relates these integrals to those in $D=4-2\eps$~\cite{Tarasov:1996br}.
Note that imposing the subgraph condition~\eqref{eq:subgraph-finiteness} on the propagator powers with $D_0=6$ requires higher powers for bubble subgraphs.
Moreover, several sectors do not contain a sufficient number of Stieltjes integrals; in these cases, some master integrals must be drawn from their super-sectors, spoiling the otherwise block-triangular structure of the differential equations.

We determine the relevant Stieltjes domain and variables in appendix~\ref{app:euclidean_ladder}, following the procedure described in section~\ref{sec:concrete-specialisations}.
The integrals are Stieltjes functions of $-x$ with threshold $R=4$,
and of $z$ with threshold $R=0$ (the limiting case $R\to0^+$ of appendix~\ref{sec:stieltjes_review}), with the other
variable held fixed in the domain $0<x<4$, $z>0$. 
Note that this domain is below the two-$W$ threshold.
It contains the points $q^2=M_Z^2$ and $q^2=M_H^2$ for the physical values of the masses.
Consequently, the 
bootstrap delivers the integrals directly at physical points without any analytic continuation.
Both variables give Hankel constraints at one and the same point $(x_0,z_0)$, from the expansion in $-x$ at fixed $z=z_0$ and from the expansion in $z$ at fixed $x=x_0$; neither variable is moved away from its physical value.

As the sample application we take the kinematic point corresponding to
the $Z$ self-energy at $q^2=M_Z^2$ with two Higgs rungs,
\begin{equation}
\begin{aligned}
  x_0=&\frac{M_Z^2}{M_W^2}=\left(\frac{911876}{803692}\right)^2
  \approx1.28734\,,
  \\
  z_0=&\frac{M_H^2}{M_W^2}=\left(\frac{1252000}{803692}\right)^2
  \approx2.42677\,,
  \end{aligned}
    \label{eq:lad2-sample-point}
\end{equation}
with $M_W=80.3692$~GeV, $M_Z=91.1876$~GeV and $M_H=125.20$~GeV~\cite{ParticleDataGroup:2024cfk}
written as exact rationals. The point lies inside the domain $0<x<4$, $z>0$ of the positivity constraints, and
all reference values below refer to it. We present results at $\eps=-1/10$, inside the admissible range
$-1/3<\eps\leq0$; the expansion in $\eps$ is obtained by repeating the evaluation at several values of $\eps$,
see section~\ref{sec:results}.

As discussed in section~\ref{sec:mi_bases}, the choice of basis has a significant impact on the size of the differential equations and hence on the cost of the bootstrap. We have two bases at our disposal, a Laporta basis ${\vect I}_{\rm L}$ and a basis ${\vect I}_{\rm S}$ of Stieltjes integrals generated as described above, each of which satisfies a system of differential equations
\begin{equation}
  \partial_u{\vect I}_{\rm B}=A_u^{({\rm B})} {\vect I}_{\rm B}\,,
  \qquad u\in\{x,z\}\,,\qquad {\rm B}={\rm L},{\rm S}\,.
  \label{eq:lad2-basis-map}
\end{equation}
We write $\vect g\equiv\vect I_{\rm S}=T\,\vect I_{\rm L}$ for the Stieltjes basis expressed in the Laporta basis. The Laporta basis, chosen following refs.~\cite{Smirnov:2020quc,Usovitsch:2020jrk}, leads to much smaller matrices, free of spurious kinematic factors and of factors mixing $\eps$ with the kinematics. We therefore use the Laporta system with the 149 scalar Stieltjes integrals of $D=6-2\eps$ as the list $\vect h=B\,\vect I_{\rm L}$ of eq.~\eqref{eq:laporta-candidate-derivatives}, with a $149\times60$ reduction matrix $B$, the Stieltjes variables $u=-x$ and $u=z$, and one known integral to fix the normalization. 

The differential equations for both the Stieltjes and Laporta bases, together with the reduction matrix $B$, are provided in the ancillary files~\cite{zenodo}. 
The matrix $T$ that maps the Stieltjes basis to the Laporta basis is obtained by selecting the corresponding rows of $B$.

\subsection{Results and comparison of numerical methods}
\label{sec:results}
We ran our Stieltjes bootstrap at the sample point~\eqref{eq:lad2-sample-point} with $\eps = -1/10$ in two different ways. We first ran the algorithm directly on the Stieltjes basis, whose differential equations are considerably more complicated. The basis contains 60 elements, one of which is normalized to 1. Table~\ref{tab:lad2-stieltjes-digits} collects the certified digits of agreement of the resulting enclosures. The number of certified digits grows by one digit per unit of $N$ across the whole range, with a spread of under one digit between the best and the worst of the bounded elements at any order. We note that the wall-clock times reported here are not monotonic in $N$. This is due to some values of $N$ producing systems with unexpectedly bad conditioning and thus requiring a retry at higher precision after failing to converge at lower precision.

As independent reference values, all 60 Laporta master integrals and all 149 Stieltjes integrals were evaluated with \textsc{AMFlow}~\cite{Liu:2017jxz,Liu:2022chg} at $\eps=-1/10$ at the sample point and at the boundary point $q^2=0$, and at further values of $\eps$ used below. The \textsc{AMFlow} runs use version 2.0 in its fixed-$\eps$ mode with $D_0=6$, the integration-by-parts reductions from its \textsc{Kira} interface (\textsc{Kira 3.1} with \textsc{Fermat}~\cite{fermat}), working precision 160 digits and the \texttt{C++} solver for the differential equations in the auxiliary mass, with all other options at their defaults. 
The 60 master integrals and the 149 Stieltjes integrals are passed directly as targets. 
All 149 Stieltjes integrals are positive at both points, as they must be, and the values satisfy the differential equation and the reduction identities to the precision of the evaluation. We checked that every enclosure returned by our bootstrap contains the corresponding \textsc{AMFlow} value.

As described before, we can also apply the bootstrap with a Laporta basis by expressing Stieltjes integrals in terms of the Laporta basis. Using all 149 Stieltjes integrals yields more accurate results at a given N , but the SDPs, with many more constraints, take much longer to solve than with the 60-element Stieltjes basis. Using only the 60 simplest Stieltjes integrals would most likely give a larger feasible set but may be much faster. We leave the optimal selection of Stieltjes integrals to use in this approach to future work.

\begin{table}
 \centering
  \footnotesize
  \begin{tabular}{rrrrr}
    \toprule
    $N$  & minimum & median & wall clock \\
    \midrule
     2 &   0.14 &  0.51 &    1.7\,min \\
     4 &   2.22 &  2.50 &    2.4\,min \\
     6 &   3.99 &  4.40 &   11.2\,min \\
     8 &  5.87 &  6.34 &    7.6\,min \\
    10 &  7.92 &  8.40 &    9.5\,min \\
    12 &  9.99 & 10.46 &    5.8\,min \\
    14 &  12.09 & 12.57 &   14.6\,min \\
    16 &  14.15 & 14.62 &   22.4\,min \\
    18 &  16.20 & 16.67 &   10.7\,min \\
    20 & 18.25 & 18.71 &   12.6\,min \\
    22 &  20.29 & 20.75 &   19.8\,min \\
    \bottomrule
  \end{tabular}
    \caption{Certified digits of the two-sided enclosures for the 59 unnormalized
    elements of the Stieltjes basis of the three-loop two-mass ladder at the
    sample point $q^2=M_Z^2$, $m_r=M_H$, $\eps=-1/10$, as a function of the
    Hankel order $N$. Digits are $-\log_{10}$ of the width of the enclosure
    relative to its midpoint. ``wall clock'' is the time to solve all 118 SDPs in the
    whole order run fully parallelized across two 64-core AMD EPYC 7702 processors. Note that this timing refers only to the time it takes to solve the SDPs once the blocks are built.} \label{tab:lad2-stieltjes-digits}
\end{table}

Applications typically need the coefficients of the expansion in $\eps$, which follow from evaluations at several
values of $\eps$ by polynomial interpolation.  We used the soft expansion and Pad\'e approximants at the 28 values $\eps=-k/160$
and $\eps=-k/640$, $k=1,\ldots,16$, from boundary values at $q^2=0$ computed with \textsc{AMFlow} to 128 digits.
For illustration, we give the first few orders of the four elements of the Stieltjes basis
that belong to the top sector,
\begin{equation}
  g_{55}=I^{(6-2\eps)}_{112211110}\,,\qquad
  g_{58}=I^{(6-2\eps)}_{111111220}\,,\qquad
  g_{59}=I^{(6-2\eps)}_{111112210}\,,\qquad
  g_{60}=I^{(6-2\eps)}_{111212110}\,.
  \label{eq:lad2-top-sector-elements}
\end{equation}
At the sample point~\eqref{eq:lad2-sample-point},
with $M_W=1$, we find
\begin{align}
  \begin{split}
    g_{55}={}&0.12010558124946322185871-0.5392847734727612746\,\eps\\
    &\hspace{-0.5cm}+2.2744725534896046\,\eps^2-8.1211156310709\,\eps^3+27.35012216684\,\eps^4+\mathcal{O}(\eps^5)\,,
  \end{split}\\
  \begin{split}
    g_{58}={}&0.037079419645972213892697-0.16875870016592120721\,\eps\\
    &\hspace{-0.5cm}+0.68197213630915128\,\eps^2-2.35546999266375\,\eps^3+7.668860953277\,\eps^4+\mathcal{O}(\eps^5)\,,
  \end{split}\\
  \begin{split}
    g_{59}={}&0.065088876705338381289527-0.29395893309090559030\,\eps\\
    &\hspace{-0.5cm}+1.21653195924519934\,\eps^2-4.28657577347636\,\eps^3+14.24693419106\,\eps^4+\mathcal{O}(\eps^5)\,,
  \end{split}\\
  \begin{split}
    g_{60}={}&0.12004904440163609872344-0.5392270877842630941\,\eps\\
    &\hspace{-0.5cm}+2.2742631457409503\,\eps^2-8.1208099253891\,\eps^3+27.34952375104\,\eps^4+\mathcal{O}(\eps^5)\,.
  \end{split}
  \label{eq:lad2-top-sector-expansions}
\end{align}

It is useful to benchmark the Stieltjes bootstrap against alternative
numerical methods on the same integrals. We compare, within
the Euclidean region, the following approaches:
\begin{itemize}
\item the Stieltjes bootstrap method developed here,
\item CM bootstrap of ref.~\cite{Ditsch:2025dhp},
\item Pad\'e approximants from the soft expansion at $q^2=0$,
\item the auxiliary-mass-flow method as implemented in \textsc{AMFlow}, and
\item the differential equation solver based on series expansions of ref.~\cite{Prisco:2025wqs} (\textsc{LINE}).
\end{itemize}
Table~\ref{tab:lad2-comparison} collects, at the sample point and $\eps=-1/10$, what each method needs,
what it delivers and what it costs, for the same 60 Stieltjes integrals. 
The times reported in Table~\ref{tab:lad2-comparison} are the wall-clock times. The Stieltjes bootstrap and CM bootstrap were run on two 64-core AMD EPYC 7702 processors, the other rows on two Xeon Silver 4216 processors. \textsc{AMFlow} was run with 24 threads, \textsc{LINE} and the Pad\'e code were run on one core. The  \textsc{LINE} transport and the soft expansion start from the same values of the 60 master integrals at $q^2=0$, computed with \textsc{AMFlow} at the same working precision as the direct evaluation. The preprocessing time of 27 min in the last two rows refers to this boundary value computation.
These comparisons illustrate both the current limitations of the Stieltjes bootstrap and its potential advantages, which we discuss in the next section.

\begin{table}[t]
  \centering
  \footnotesize
  \begin{tabular}{>{\raggedright\arraybackslash}p{0.19\textwidth}>{\raggedright\arraybackslash}p{0.26\textwidth}>{\raggedright\arraybackslash}p{0.12\textwidth}>{\raggedright\arraybackslash}p{0.31\textwidth}}
    \specialrule{1.2pt}{0pt}{2pt}
    method & input & output & cost \\
    \specialrule{1.2pt}{2pt}{2pt}
    Stieltjes bootstrap (this work) & DE, one normalization & 20 digits & 20\,min solve,\newline plus 12\,min preprocessing \\ \cmidrule(lr){1-4}
    Complete-monotonicity bootstrap & DE, one normalization & 10 digits & 28\,min solve,\newline plus 13\,min preprocessing \\ \cmidrule(lr){1-4}
    AMFlow (direct) & family definition only & 80 digits & 24\,min \\ \cmidrule(lr){1-4}
    LINE transport & DE, 60 values at $q^2=0$ & 56 digits & 26\,s,\newline plus 27\,min (boundary, AMFlow) \\ \cmidrule(lr){1-4}
    Soft expansion + Pad\'e (this work) & DE, $T$, 60 values at $q^2=0$ & 24 digits & 44\,s,\newline plus 27\,min (boundary, AMFlow) \\
    \specialrule{1.2pt}{2pt}{0pt}
  \end{tabular}
%    \caption{Comparison of the methods on the three-loop family.}
\caption{Input, achieved precision and wall-clock cost of the five methods
for the 60 Stieltjes integrals of the three-loop two-mass family, at the
sample point and $\epsilon=-1/10$. % Preprocessing is listed separately; for \textsc{LINE} and Padé it is the computation of the boundary values at $q^2=0$ with \textsc{AMFlow}.
}
  \label{tab:lad2-comparison}
\end{table}

%=====================================================================
\section{Summary and discussion}
\label{sec:summary}
%=====================================================================

We proved that scalar Feynman integrals 
are Stieltjes functions in each of the kinematic variables~\eqref{eq:stieltjes_vars}, for arbitrary connected graphs,
via the manifestly non-negative Symanzik decomposition~\eqref{eq:main_decomp}. We showed how to build master-integral bases that are
entirely Stieltjes, turned the property into rigorous
Hankel/SDP bounds, and applied it to a state-of-the-art three-loop integral family.

It is fair to ask how the bootstrap compares with existing numerical methods. 
Table~\ref{tab:lad2-comparison} shows that for the three-loop family the Stieltjes bootstrap is not yet competitive with \textsc{AMFlow} as a way of getting numbers but is comparable to other methods in terms of time and accuracy. \textsc{AMFlow} evaluates the same integrals to more digits in less time, and once the values
at $q^2=0$ are known, transporting them is a matter of seconds. However, the bootstrap introduced here still has many avenues for improvement and offers distinct advantages. It
uses nothing but the differential equation and one normalization, and it returns certified bounds such that the true value of the integrals is guaranteed to lie within the intervals reported. Moreover, the bootstrap can be used to obtain a value for a single integral in significantly less time, since only two SDPs need to be solved. For instance, on the state-of-the-art example discussed in section \ref{sec:examples}, the bootstrap could be used to produce values for the most difficult integrals in at most 5 minutes of SDP solve time plus 12 minutes of preprocessing.

The conceptually most direct way to formulate our method is by writing a system of differential equations for a basis of Stieltjes master integrals. As we commented in sections~\ref{sec:diff-eq-sdp} and~\ref{sec:examples}, one may also use any convenient basis, for example a Laporta basis, to derive differential equations, and supply the information about Stieltjes integrals as additional reduction equations in this basis. This has the advantage that the differential equation matrix is smaller to store, and only one additional integral reduction is needed compared with a standard differential-equation computation. We have not yet explored this approach fully but we believe that it may allow for both smaller preprocessing times and simpler SDPs once fully optimized.

The bounds converge exponentially in the truncation order for the examples of this paper. As discussed in section~\ref{sec:convergence}, whether the convergence is exponential, algebraic or absent appears to be governed by the singularities of the homogeneous solutions of the differential equations that positivity has to exclude. A general proof of this picture, and an a posteriori certification of the bounds against the numerical errors of the semidefinite solver, are left for future work. A deeper understanding of the convergence behavior may also have implications for the runtime of our method. Given a precise rate of convergence in terms of $N$, the integral family, and the differential equation, one could predict the width at any $N$. A single feasible point of the SDP at that $N$ would then approximate all
coupled integrals to known accuracy, requiring one SDP instead of $2(n{-}1)$.

The use of Pad\'e approximants for the evaluation of Feynman integrals has a long and productive history, with early work demonstrating their remarkable accuracy when combined with Taylor expansions and conformal mappings \cite{Broadhurst:1993mw,Fleischer:1994ef}, and subsequent applications extending this approach to multi-loop calculations \cite{Fael:2021xdp,Davies:2023vmj}.
In these applications, Padé approximants are typically used heuristically: spurious poles are discarded and the spread among approximants is taken as an empirical error estimate. While effective in practice, this procedure offers no general guarantee of convergence or rigorous error control. The Stieltjes property provides such a theoretical foundation, as was already noted in ref.~\cite{Broadhurst:1993mw}. For Stieltjes functions, diagonal Padé approximants are guaranteed to converge, their poles lie on the physical cut, and neighboring Padé sequences provide rigorous upper and lower bounds~\cite{Baker:1996,Cuyt:2008}.
Since Feynman integrals can be decomposed, via integration-by-parts reduction, into a basis of Stieltjes master integrals, as shown here, this explains the empirical success of Padé methods while turning their error estimates into rigorous bounds.
We also note that this method already yields certified results within seconds; if the boundary values were obtained analytically---which is within reach for vacuum integrals---the method would be quite competitive for the evaluation of Feynman integrals.

In this paper, we evaluated the integrals at fixed negative values of
$\eps$. Applications to scattering amplitudes usually require the
coefficients of their Laurent expansion around $\eps=0$.
Extrapolating such expansions from numerical information at finite $\eps$ is routinely done by repeating the numerical
evaluation at several values of $\eps$ and reconstructing the
Laurent expansion~\cite{Liu:2022chg, Zeng:2023jek}.
Alternatively, the infrared- and ultraviolet-finite remainders entering cross-section calculations can be obtained directly from amplitudes evaluated at fixed values of $\eps$~\cite{Bi:2023bnq,Becchetti:2026yxl}.

An important follow-up direction concerns the analytic continuation of the results to
physical scattering kinematics.
While the Stieltjes bootstrap and Padé interpolations accurately describe the functions in the cut complex plane, convergence is slow near the branch cuts. Analytic continuation therefore requires a dedicated method and implementation, which is work in progress.

\section*{Acknowledgements}

J.M.H.\ and Q.Y.\ thank the participants of the conference Loops and Legs in Quantum Field Theory 2026 (Bayreuth) for valuable discussions on the use of Pad\'e approximants in high energy physics.
We are grateful to the Max Planck--IAS--NTU Center for Particle Physics, Cosmology, and Geometry for support.
Funded by the European Union (ERC, UNIVERSE PLUS, 101118787). Views and opinions expressed are however those of the authors only and do not necessarily reflect those of the European Union or the European Research Council Executive Agency. Neither the European Union nor the granting authority can be held responsible for them.
S.Z.\ was supported by the Swiss National Science Foundation (SNSF) under the Ambizione grant No.~215960. P.R.\ receives support from the Max Planck--IAS--NTU Center, partly funded by NSTC Grant No.~114-2923-M-002-011-MY5.

\appendix

%=====================================================================

\section{Stieltjes functions and Pad\'e approximants}
\label{sec:stieltjes_review}
%=====================================================================
In this appendix, we collect the relevant background on Stieltjes functions and Pad\'e approximation. Standard references
include refs.~\cite{Widder:1941,Baker:1996,Cuyt:2008}.
The Pad\'e bounds~\eqref{eq:pade_bracketing} are used in section~\ref{sec:massive-box}, and the moment expansion~\eqref{eq:moment_expansion} underlies the Hankel constraints of section~\ref{app:hankel_stieltjes}.

A real-valued function $f$ on $(-R,\infty)$, with $R>0$, is called a {\it Stieltjes function}
with threshold $R$ if it admits the representation
\begin{equation}
  f(x)=\int_R^\infty\frac{\dd\mu(t)}{x+t}\,,
  \qquad \dd\mu(t)\geq0\,.
  \label{eq:stieltjes_R}
\end{equation}
Here $\dd\mu(t)$ may include both ordinary and delta functions.
The real domain of $f$ is $x>-R$, where the integral is finite. Non-negative linear combinations of Stieltjes functions are again
Stieltjes functions. The representation in eq.~\eqref{eq:stieltjes_R} also defines an analytic
function in the complex plane with the cut $(-\infty,-R]$ removed. It maps
the upper half-plane to the lower half-plane, i.e.
\begin{equation}
  \operatorname{Im}f(z)\leq0
  \qquad \text{for }\operatorname{Im}z>0\,.
  \label{eq:stieltjes_pick}
\end{equation}
The discontinuity across the cut determines the integration weight. In cases where it can be written as $\dd\mu(t)=\rho(t)\,\dd t$ with $\rho(t)$ an ordinary
function, we have
\begin{equation}
  \rho(t)=-\frac{1}{\pi}\operatorname{Im}f(-t+\ii 0^+)\,.
  \label{eq:stieltjes_density}
\end{equation}
Since $R>0$, $f$ is analytic at $x=0$, and its Taylor coefficients are moments of a positive measure,
\begin{equation}
  f(x)=\sum_{k=0}^\infty(-1)^k c_k\,x^k\,,
  \qquad
  c_k=\int_R^\infty\frac{\dd\mu(t)}{t^{k+1}}=\int_0^{1/R}u^k\,\dd\nu(u)\,,
  \label{eq:moment_expansion}
\end{equation}
where $\dd\nu(u)=u\,\dd\mu(1/u)\geq0$ is supported on $[0,1/R]$. All $c_k$ are finite, and the series converges for $|x|<R$.
We denote the {\it Pad\'e approximant} of the series~\eqref{eq:moment_expansion} by
\begin{equation}
    [L/M]_f(x) \coloneq \frac{P_L(x)}{Q_M(x)} \,.
\end{equation}
It is a ratio of two
polynomials $P_L$ and $Q_M$, with $\deg P_L\leq L$, $\deg Q_M\leq M$, and $Q_M(0)=1$. The two polynomials are determined by asking that they agree with the series expansion of $f$ through order $x^{L+M}$, i.e., by requiring that
\begin{equation}
  Q_M(x)f(x)-P_L(x)=O\!\left(x^{L+M+1}\right)\,.
  \label{eq:pade_definition}
\end{equation}
Let us denote the two Pad\'e approximant sequences by
\begin{equation}
  \mathcal{P}_N^-(x)=[N-1/N]_f(x)\,,
  \qquad
  \mathcal{P}_N^+(x)=[N/N]_f(x)\,.
  \label{eq:pade_shifted_sequences}
\end{equation}
The following classical results hold~\cite{Baker:1996}:
\begin{itemize}
 \item[(a)] \textit{Bounds on the real axis.}
For $x\geq0$, the subdiagonal approximants give increasing
lower bounds, while the diagonal approximants give decreasing
upper bounds:
\begin{equation}
\mathcal{P}_N^-(x)
\leq \mathcal{P}_{N+1}^-(x)
\leq f(x)
\leq \mathcal{P}_{N+1}^+(x)
\leq \mathcal{P}_N^+(x) \,.
\label{eq:pade_bracketing}
\end{equation}
For $-R<x<0$, both sequences instead give lower bounds,
with the diagonal approximant at least as accurate as the
subdiagonal approximant:
\begin{equation}
\mathcal{P}_N^-(x)\leq\mathcal{P}_N^+(x)\leq f(x).
\label{eq:pade_negative_bounds}
\end{equation}
More precisely, their errors satisfy~\cite{Gilewicz:2010}
\begin{equation}
0\leq f(x)-\mathcal{P}_N^+(x)
\leq \frac{|x|}{R}
\left(f(x)-\mathcal{P}_N^-(x)\right)\,.
\label{eq:pade_negative_error}
\end{equation}
Thus, the superscripts $\pm$ indicate upper and lower bounds
only for $x\geq0$. At $x=0$, all approximants equal $f(0)$.
\item[(b)] \textit{Pole location.}
  All poles of $\mathcal{P}_N^-$ and $\mathcal{P}_N^+$ lie on
  $(-\infty,-R]$ and have positive residues.
\item[(c)] \textit{Convergence.}
Since the $c_k$ are moments of a measure with compact support, they determine $f$ uniquely, and both Pad\'e sequences converge to $f$ throughout the cut plane, uniformly on closed bounded regions that stay away from the cut, in particular for all $x>-R$, beyond the radius of convergence of the series.
\end{itemize}

\section{Connection matrices for the one-loop box integral family}
\label{app:matrices}

The three connection matrices of the Stieltjes basis of the one-loop box family discussed in section~\ref{sec:massive-box} are given by
\begingroup
\setlength{\arraycolsep}{3pt}
\renewcommand{\arraystretch}{1.2}
\setbox0=\hbox{$\displaystyle
A_{m^2}
=
\begin{pmatrix}
-\dfrac{2(1-2\eps)(s+t)}{\Delta} & -\dfrac{2s}{\Delta} & -\dfrac{2t}{\Delta} & 0 & 0 & 0 \\
0 & -\dfrac{\eps}{m^2} & 0 & \dfrac{1}{2m^2} & 0 & 0 \\
0 & 0 & -\dfrac{\eps}{m^2} & 0 & \dfrac{1}{2m^2} & 0 \\
0 & 0 & 0 & -\dfrac{2(1+2\eps)}{4m^2-s} & 0 & \dfrac{4}{4m^2-s} \\
0 & 0 & 0 & 0 & -\dfrac{2(1+2\eps)}{4m^2-t} & \dfrac{4}{4m^2-t} \\
0 & 0 & 0 & 0 & 0 & -\dfrac{1+\eps}{m^2}
\end{pmatrix},
$}
\setbox2=\hbox{$\displaystyle
A_s
=
\begin{pmatrix}
\dfrac{2m^2(s+t)(2s+t)-st(s+\eps t)}{s(s+t)\Delta} & \dfrac{2m^2t}{(s+t)\Delta} & \dfrac{2m^2t^2}{s(s+t)\Delta} & \dfrac{s-4m^2}{2(1-2\eps)s(s+t)} & \dfrac{4m^2-t}{2(1-2\eps)s(s+t)} & 0 \\
0 & -\dfrac{1}{s} & 0 & -\dfrac{1}{2s} & 0 & 0 \\
0 & 0 & 0 & 0 & 0 & 0 \\
0 & 0 & 0 & \dfrac{2m^2-(1+\eps)s}{s(s-4m^2)} & 0 & \dfrac{4m^2}{s(s-4m^2)} \\
0 & 0 & 0 & 0 & 0 & 0 \\
0 & 0 & 0 & 0 & 0 & 0
\end{pmatrix},
$}
\setbox4=\hbox{$\displaystyle
A_t
=
\begin{pmatrix}
\dfrac{2m^2(s+t)(s+2t)-st(\eps s+t)}{t(s+t)\Delta} & \dfrac{2m^2s^2}{t(s+t)\Delta} & \dfrac{2m^2s}{(s+t)\Delta} & \dfrac{4m^2-s}{2(1-2\eps)t(s+t)} & \dfrac{t-4m^2}{2(1-2\eps)t(s+t)} & 0 \\
0 & 0 & 0 & 0 & 0 & 0 \\
0 & 0 & -\dfrac{1}{t} & 0 & -\dfrac{1}{2t} & 0 \\
0 & 0 & 0 & 0 & 0 & 0 \\
0 & 0 & 0 & 0 & \dfrac{2m^2-(1+\eps)t}{t(t-4m^2)} & \dfrac{4m^2}{t(t-4m^2)} \\
0 & 0 & 0 & 0 & 0 & 0
\end{pmatrix},
$}
\dimen0=\wd0
\ifdim\wd2>\dimen0 \dimen0=\wd2\fi
\ifdim\wd4>\dimen0 \dimen0=\wd4\fi
\[
\resizebox{0.98\textwidth}{!}{\makebox[\dimen0][c]{\copy0}}
\]
\[
\resizebox{0.98\textwidth}{!}{\makebox[\dimen0][c]{\copy2}}
\]
\[
\resizebox{0.98\textwidth}{!}{\makebox[\dimen0][c]{\copy4}}
\]
\endgroup
where 
\begin{equation}
 \Delta=st-4m^2(s+t)\,.
  \label{eq:box_delta}
  \end{equation}
The block-triangular structure reflects the ordering of the sectors.

\section{Stieltjes domain of the two-mass three-loop ladder integrals}
\label{app:euclidean_ladder}

In order to determine the
relevant Stieltjes kinematic domain
of the two-mass three-loop ladder integrals discussed in section~\ref{sec:examples}, we proceed as explained in section~\ref{sec:proof} from the positivity of the
second Symanzik polynomial.
With the overall mass scale
removed, it reads
\begin{equation}
  \cF
  =\cU\sum_{i=1}^{6}\alpha_i
   +z\,\cU(\alpha_7+\alpha_8)
   -x\,\cF_0\,,
  \label{eq:lad2-symanzik}
\end{equation}
where $\cF_0$ is the sum of the two-forest monomials
separating the external vertices. Since $\cU$ and
$\cF_0$ have non-negative coefficients, this expression
is manifestly positive in the Euclidean region
$x<0$, $z>0$. However, the positivity domain actually extends to
$x<4$. To show this, define
\begin{equation}
  \mathcal{Q}=\cU\sum_{i=1}^{6}\alpha_i-4 \, \cF_0\,.
  \label{eq:lad2-positive-polynomial}
\end{equation}
The quadratic form in the three loop momenta gives
the matrix
\begin{equation}
  \mathsf M=
  \begin{pmatrix}
    \alpha_1+\alpha_4+\alpha_7
      &-\alpha_7&0\\
    -\alpha_7
      &\alpha_2+\alpha_5+\alpha_7+\alpha_8
      &-\alpha_8\\
    0&-\alpha_8
      &\alpha_3+\alpha_6+\alpha_8
  \end{pmatrix}\,,
  \qquad \cU=\det\mathsf M\,.
  \label{eq:lad2-loop-matrix}
\end{equation}
Completing the square in the loop momenta yields
\begin{equation}
  \mathcal{Q}=\vect d^{\,T}\operatorname{adj}(\mathsf M) \, \vect d\,,
  \qquad
  \vect d=
  \begin{pmatrix}
    \alpha_1-\alpha_4\\
    \alpha_2-\alpha_5\\
    \alpha_3-\alpha_6
  \end{pmatrix}\,,
  \label{eq:lad2-positivity}
\end{equation}
where $\operatorname{adj}(\mathsf M)$ is the adjugate
matrix. For positive Feynman parameters, $\mathsf M$
is positive definite, and hence
$\operatorname{adj}(\mathsf M)=\det(\mathsf M)\mathsf M^{-1}$
is positive definite. It follows that $\mathcal{Q}\geq0$,
including at non-negative Feynman parameters by continuity.
We can therefore rewrite
eq.~\eqref{eq:lad2-symanzik} as
\begin{equation}
  \cF
  =\mathcal{Q}+(4-x)\cF_0
   +z\,\cU(\alpha_7+\alpha_8)\,.
  \label{eq:lad2-threshold}
\end{equation}
All three terms are non-negative for $x<4$ and $z>0$.
Thus $4-x$ and $z$ are adapted positive coordinates,
as in section~\ref{sec:concrete-specialisations}.

\bibliographystyle{JHEP}
\bibliography{refs}

\end{document}